\input harvmac.tex
\input amssym.tex


\font\teneurm=eurm10 \font\seveneurm=eurm7 \font\fiveeurm=eurm5

\newfam\eurmfam

\textfont\eurmfam=\teneurm \scriptfont\eurmfam=\seveneurm

\scriptscriptfont\eurmfam=\fiveeurm

\def\eurm#1{{\fam\eurmfam\relax#1}}

 \font\teneusm=eusm10 \font\seveneusm=eusm7 \font\fiveeusm=eusm5

\newfam\eusmfam

\textfont\eusmfam=\teneusm \scriptfont\eusmfam=\seveneusm

\scriptscriptfont\eusmfam=\fiveeusm

\def\eusm#1{{\fam\eusmfam\relax#1}}

\font\tencmmib=cmmib10 \skewchar\tencmmib='177

\font\sevencmmib=cmmib7 \skewchar\sevencmmib='177

\font\fivecmmib=cmmib5 \skewchar\fivecmmib='177

\newfam\cmmibfam

\textfont\cmmibfam=\tencmmib \scriptfont\cmmibfam=\sevencmmib

\scriptscriptfont\cmmibfam=\fivecmmib




\lref\AtiyahF{M.F.~Atiyah, ``Floer homology,'' Progr. Math.
Birkhauser 133 (1995) 105.}

\lref\Atiyah{M.F.~Atiyah, ``The Geometry and Physics of Knots,''
Cambridge Univ. Press, 1990.}

\lref\Rasmussen{J.~Rasmussen, ``Floer homology and knot
complements," math.GT/0306378.}

\lref\OShf{P.~Ozsvath, Z.~Szabo, ``Holomorphic disks and
topological invariants for closed three-manifolds,''
Ann. of Math. {\bf 159} (2004) 1027, math.SG/0101206.}

\lref\OShfk{P.~Ozsvath, Z.~Szabo, ``Holomorphic disks and knot
invariants," Adv. Math. {\bf 186} (2004) 58, math.GT/0209056.}

\lref\OSlens{P.~Ozsvath, Z.~Szabo, ``On knot Floer homology and
lens space surgeries,'' math.GT/0303017.}

\lref\OSappl{P.~Ozsvath, Z.~Szabo, ``Holomorphic disks and
three-manifold invariants: properties and applications,''
Ann. of Math. {\bf 159} (2004) 1159, math.SG/0105202.}

\lref\OSreview{P.~Ozsvath, Z.~Szabo, ``Heegaard diagrams and
holomorphic disks,'' {\it Different faces of geometry},
Int. Math. Ser. (N. Y.), 3, Kluwer/Plenum, New York (2004) 301, math.GT/0403029.}

\lref\OShfl{P.~Ozsvath, Z.~Szabo, ``Holomorphic disks and link
invariants,'' math.GT/0512286.}

\lref\MengT{G.~Meng, C.~Taubes, ``$SW=$ Milnor Torsion,'' Math.
Res. Lett. {\bf 3} (1996) 661.}

\lref\GSV{S.~Gukov, A.~Schwarz and C.~Vafa, ``Khovanov-Rozansky
homology and topological strings,''
Lett. Math. Phys. {\bf 74} (2005) 53, hep-th/0412243.}

\lref\WittenJones{ E.~Witten, ``Quantum Field Theory And The Jones
Polynomial,'' Commun.\ Math.\ Phys.\  {\bf 121}, 351 (1989).}

\lref\HOMFLY{P.~Freyd, D.~Yetter, J.~Hoste, W.~Lickorish,
K.~Millett, A.~Oceanu, ``A New Polynomial Invariant of Knots and
Links,'' Bull. Amer. Math. Soc. {\bf 12} (1985) 239.}

\lref\Wittencsstring{ E.~Witten, ``Chern-Simons gauge theory as a
string theory,'' Prog.\ Math.\  {\bf 133} (1995) 637,
hep-th/9207094.}

\lref\OV{ H.~Ooguri, C.~Vafa, ``Knot Invariants and Topological
Strings,'' Nucl.Phys. {\bf B577} (2000) 419, hep-th/9912123.}

\lref\Khovanov{M.~ Khovanov, ``A categorification of the Jones
polynomial,''  Duke Math. J. {\bf 101} (2000) 359, math.QA/9908171.}

\lref\Khovanovii{M.~ Khovanov, ``Categorifications of the colored
Jones polynomial,''  J. Knot Theory Ramifications {\bf 14} (2005) 111, math.QA/0302060.}

\lref\Khovanoviii{M.~ Khovanov, ``$sl(3)$ link homology I,''
Algebr. Geom. Topol. {\bf 4} (2004) 1045, math.QA/0304375.}

\lref\Khovanoviv{M.~ Khovanov, ``An invariant of tangle cobordisms,''
Trans. Amer. Math. Soc. {\bf 358}  (2006) 315, math.QA/0207264.}

\lref\Khovanovtangles{M.~Khovanov, ``A functor-valued invariant of
tangles,'' Algebr. Geom. Topol. {\bf 2} (2002) 665, math.QA/0103190.}

\lref\DBN{D.~Bar-Natan, ``On Khovanov's categorification of the
Jones polynomial,'' Algebr. Geom. Topol. {\bf 2} (2002) 337, math.QA/0201043.}

\lref\DBNnews{D.~Bar-Natan, ``Some Khovanov-Rozansky
Computations'' \semi
{http://www.math.toronto.edu/~drorbn/Misc/KhovanovRozansky/index.html}}

\lref\Jacobsson{M.~Jacobsson, ``An invariant of link cobordisms from
Khovanov homology,'' Algebr. Geom. Topol. {\bf 4} (2004) 1211, math.GT/0206303.}

\lref\RKhovanov{M.~Khovanov, L.~Rozansky, ``Matrix factorizations
and link homology,'' math.QA/0401268.}

\lref\RKhovanovii{M.~Khovanov, L.~Rozansky, ``Matrix factorizations
and link homology II,'' math.QA/0505056.}

\lref\Rfoam{L.~Rozansky, ``Topological A-models on seamed Riemann
surfaces,'' hep-th/0305205.}

\lref\KRfoam{M.~Khovanov and L.~Rozansky, ``Topological
Landau-Ginzburg models on a world-sheet foam,'' hep-th/0404189.}

\lref\GopakumarV{R.~Gopakumar and C.~Vafa, ``On the gauge
theory/geometry correspondence,'' Adv.\ Theor.\ Math.\ Phys.\
{\bf 3} (1999) 1415, hep-th/9811131.}

\lref\GViii{R.~Gopakumar and C.~Vafa, ``M-theory and topological
strings. I,II,'' hep-th/9809187; hep-th/9812127.}

\lref\KKV{S.~Katz, A.~Klemm and C.~Vafa, ``M-theory, topological
strings and spinning black holes,'' Adv.\ Theor.\ Math.\ Phys.\
{\bf 3} (1999) 1445, hep-th/9910181.}

\lref\mirbook{``Mirror Symmetry'' (Clay Mathematics Monographs, V.
1), K.~Hori et.al. ed, American Mathematical Society, 2003.}

\lref\HV{K.~Hori, C.~Vafa, ``Mirror Symmetry,'' hep-th/0002222;
K.~Hori, A.~Iqbal, C.~Vafa, ``D-Branes And Mirror Symmetry,''
hep-th/0005247.}

\lref\AKV{M.~Aganagic, A.~Klemm, C.~Vafa, ``Disk Instantons,
Mirror Symmetry and the Duality Web,'' hep-th/0105045.}

\lref\AKMV{M.~Aganagic, A.~Klemm, M.~Marino and C.~Vafa, ``Matrix
model as a mirror of Chern-Simons theory,'' JHEP {\bf 0402}, 010
(2004), hep-th/0211098.}

\lref\AAHV{B.~Acharya, M.~Aganagic, K.~Hori and C.~Vafa,
``Orientifolds, mirror symmetry and superpotentials,''
hep-th/0202208.}

\lref\LMV{J.~M.~F.~Labastida, M.~Marino and C.~Vafa, ``Knots,
links and branes at large N,'' JHEP {\bf 0011}, 007 (2000),
hep-th/0010102.}

\lref\LMtorus{J.~M.~F.~Labastida and M.~Marino, ``Polynomial
invariants for torus knots and topological strings,'' Commun.\
Math.\ Phys.\  {\bf 217} (2001) 423, hep-th/0004196.}

\lref\LMqa{J.~M.~F.~Labastida and M.~Marino, ``A new point of view
in the theory of knot and link invariants,'' math.qa/0104180.}

\lref\HSTa{S.~Hosono, M.-H.~Saito, A.~Takahashi, ``Holomorphic
Anomaly Equation and BPS State Counting of Rational Elliptic
Surface,'' Adv.Theor.Math.Phys. {\bf 3} (1999) 177.}

\lref\HSTb{S.~Hosono, M.-H.~Saito, A.~Takahashi, ``Relative
Lefschetz Action and BPS State Counting,'' Internat. Math. Res.
Notices, (2001), No. 15, 783.}

\lref\Kprivate{M.~Khovanov, private communication.}

\lref\Taubes{ C.~Taubes, ``Lagrangians for the Gopakumar-Vafa
conjecture,'' math.DG/0201219.}

\lref\Wittenams{E.~Witten, ``Dynamics of Quantum Field Theory,''
{\it Quantum Fields and Strings: A Course for Mathematicians} (P.
Deligne, {\it et.al.} eds.), vol. 2, AMS Providence, RI, (1999)
pp. 1313-1325.}

\lref\LVW{W.~Lerche, C.~Vafa and N.~P.~Warner, ``Chiral Rings In
N=2 Superconformal Theories,'' Nucl.\ Phys.\ B {\bf 324}, 427
(1989).}

\lref\HMoore{J.~A.~Harvey and G.~W.~Moore, ``On the algebras of
BPS states,'' Commun.\ Math.\ Phys.\  {\bf 197} (1998) 489,
hep-th/9609017.}

\lref\IqbalV{T.~J.~Hollowood, A.~Iqbal and C.~Vafa, ``Matrix
models, geometric engineering and elliptic genera,''
hep-th/0310272.}

\lref\Schwarz{A.~Schwarz, ``New topological invariants arising in
the theory of quantized fields,'' Baku International Topological
Conf., Abstracts (part II) (1987).}

\lref\SchwarzS{A.~Schwarz and I.~Shapiro, ``Some remarks on
Gopakumar-Vafa invariants,'' hep-th/0412119.}

\lref\Aspinwallrev{ P.~S.~Aspinwall, ``D-branes on Calabi-Yau
manifolds,'' hep-th/0403166.}

\lref\MNOP{D.~Maulik, N.~Nekrasov, A.~Okounkov, R.~Pandharipande,
``Gromov-Witten theory and Donaldson-Thomas theory, I,''
math.AG/0312059.}

\lref\Katz{S.~Katz, ``Gromov-Witten, Gopakumar-Vafa, and
Donaldson-Thomas invariants of Calabi-Yau threefolds,''
math.ag/0408266.}

\lref\FultonH{W.~Fulton, J.~Harris, ``Representation Theory: A
First Course,'' Springer-Verlag 1991.}

\lref\Kontsevich{M.~Kontsevich, unpublished.}

\lref\KapustinLi{A.~Kapustin and Y.~Li, ``D-branes in
Landau-Ginzburg models and algebraic geometry,'' JHEP {\bf 0312}
(2003) 005, hep-th/0210296.}

\lref\KapustinLii{A.~Kapustin and Y.~Li, ``Topological correlators
in Landau-Ginzburg models with boundaries,'' Adv.\ Theor.\ Math.\
Phys.\  {\bf 7} (2004) 727, hep-th/0305136.}

\lref\KapustinLiii{A.~Kapustin and Y.~Li, ``D-branes in
topological minimal models: The Landau-Ginzburg approach,'' JHEP
{\bf 0407} (2004) 045, hep-th/0306001.}

\lref\Brunneriii{ I.~Brunner, M.~Herbst, W.~Lerche and J.~Walcher,
``Matrix factorizations and mirror symmetry: The cubic curve,''
hep-th/0408243.}

\lref\Brunner{I.~Brunner, M.~Herbst, W.~Lerche and B.~Scheuner,
``Landau-Ginzburg realization of open string TFT,''
hep-th/0305133.}

\lref\HLLii{M.~Herbst, C.~I.~Lazaroiu and W.~Lerche, ``D-brane
effective action and tachyon condensation in topological minimal
models,'' hep-th/0405138.}

\lref\HLL{M.~Herbst, C.~I.~Lazaroiu and W.~Lerche,
``Superpotentials, A(infinity) relations and WDVV equations for
open topological strings,'' hep-th/0402110. }

\lref\LercheJW{W.~Lerche and J.~Walcher, ``Boundary rings and N =
2 coset models,'' Nucl.\ Phys.\ B {\bf 625} (2002) 97,
hep-th/0011107.}

\lref\HoriJW{K.~Hori and J.~Walcher, ``F-term equations near
Gepner points,'' hep-th/0404196.}

\lref\Orlov{D.~Orlov, ``Triangulated Categories of Singularities
and D-Branes in Landau-Ginzburg Orbifold,'' math.AG/0302304.}

\lref\Emanuelii{S.~K.~Ashok, E.~Dell'Aquila, D.~E.~Diaconescu and
B.~Florea, ``Obstructed D-branes in Landau-Ginzburg orbifolds,''
hep-th/0404167.}

\lref\Emanuel{ S.~K.~Ashok, E.~Dell'Aquila and D.~E.~Diaconescu,
``Fractional branes in Landau-Ginzburg orbifolds,''
hep-th/0401135.}

\lref\Guadagnini{E~.Guadagnini, ``The Link Invariants of the
Chern-Simons Field Theory: New Developments in Topological Quantum
Field Theory,'' Walter de Gruyter Inc., 1997.}

\lref\MOY{H.~Murakami, T.~Ohtsuki, S.~Yamada, ``HOMFLY polynomial
via an invariant of colored plane graphs,'' Enseign. Math. {\bf
44} (1998) 325.}

\lref\Shumakovitch{A.~Shumakovitch, {\it KhoHo} --- a program for
computing and studying Khovanov homology,
{http://www.geometrie.ch/KhoHo}}

\lref\DBNtangles{D.~Bar-Natan, ``Khovanov's Homology for Tangles
and Cobordisms,'' Geom. Topol. {\bf 9} (2005) 1443, math.GT/0410495.}

\lref\Lee{E.S.~Lee, ``The support of the Khovanov's invariants for
alternating knots,'' math.GT/0201105.}

\lref\HTi{A.~Hanany and D.~Tong,``Vortices, instantons and
branes,'' JHEP {\bf 0307}, 037 (2003), hep-th/0306150.}

\lref\HTii{A.~Hanany and D.~Tong, ``Vortex strings and
four-dimensional gauge dynamics,'' JHEP {\bf 0404}, 066 (2004),
hep-th/0403158.}

\lref\Bradlowi{S.~Bradlow, ``Vortices in Holomorphic Line Bundles
over Closed Kahler Manifolds,'' Commun. Math. Phys. {\bf 135}
(1990) 1.}

\lref\BGPvortices{S.~Bradlow, O.~Garcia-Prada, ``Non-abelian
monopoles and vortices,'' math.AG/9602010.}

\lref\OTi{C.~Okonek, A.~Teleman, ``The Coupled Seiberg-Witten
Equations, vortices, and Moduli spaces of stable pairs,''
alg-geom/9505012.}

\lref\OTii{C.~Okonek, A.~Teleman, ``Quaternionic Monopoles,''
alg-geom/9505029.}

\lref\OTiii{C.~Okonek, A.~Teleman, ``Recent Developments in
Seiberg-Witten Theory and Complex Geometry,'' alg-geom/9612015.}

\lref\Gornik{B.~Gornik, ``Note on {K}hovanov link cohomology,''
math.QA/0402266.}

\lref\DGR{ N.~Dunfield, S.~Gukov, J.~Rasmussen, ``The Superpolynomial for
Knot Homologies,'' Experiment. Math. {\bf 15} (2006) 129, math.GT/0505662.}

\lref\Fukaya{K.~Fukaya, ``Floer homology for 3-manifolds with
boundary,'' in {\it Topology, Geometry, and Field Theory}, World
Sci. Publishing, River Edge, NJ, 1994.}

\lref\KMbook{P.B.~Kronheimer, T.S.~Mrowka, ``Floer homology for
Seiberg-Witten monopoles,'' in preparation.}

\lref\HKhovanov{R.~Huerfano, M.~Khovanov, ``A category for the
adjoint representation,'' J. Algebra {\bf 246} (2001) 514, math.QA/0002060.}

\lref\SeidelSmith{P.~Seidel, I.~Smith, ``A link invariant from the
symplectic geometry of nilpotent slices,'' math.SG/0405089.}

\lref\CFrenkel{L.~Crane, I.~Frenkel, ``Four-dimensional
topological quantum field theory, Hopf categories, and the
canonical bases,'' J. Math. Phys. {\bf 35} (1994) 5136, hep-th/9405183.}

\lref\Fischer{J.~Fischer, ``2-categories and 2-knots,'' Duke Math.
J. {\bf 75} (1994) 493.}

\lref\Seibergetal{P.~C.~Argyres, M.~R.~Plesser and N.~Seiberg,
``The Moduli Space of N=2 SUSY {QCD} and Duality in N=1 SUSY
{QCD},'' Nucl.\ Phys.\ B {\bf 471} (1996) 159, hep-th/9603042.}

\lref\DGNV{R.~Dijkgraaf, S.~Gukov, A.~Neitzke and C.~Vafa,
``Topological M-theory as unification of form theories of
gravity,'' hep-th/0411073.}

\lref\JaffeT{A.~Jaffe, C.~Taubes, ``Vortices and monopoles,''
Birkhauser, Boston MA, 1980.}

\lref\HoriH{A.~Hanany and K.~Hori, ``Branes and N = 2 theories in
two dimensions,'' Nucl.\ Phys.\ B {\bf 513} (1998) 119,
hep-th/9707192.}

\lref\WittenDonaldson{E.~Witten, ``Topological Quantum Field
Theory,'' Commun.\ Math.\ Phys.\  {\bf 117} (1988) 353.}

\lref\Wittenmonopoles{E.~Witten, ``Monopoles and four manifolds,''
Math.\ Res.\ Lett.\  {\bf 1} (1994) 769, hep-th/9411102.}

\lref\Wittensigma{E.~Witten, ``Topological Sigma Models,''
Commun.\ Math.\ Phys.\  {\bf 118} (1988) 411.}

\lref\Wittenmirror{E.~Witten, ``Mirror manifolds and topological
field theory,'' hep-th/9112056.}

\lref\BaezD{J.~C.~Baez and J.~Dolan, ``Higher dimensional algebra
and topological quantum field theory,'' J.\ Math.\ Phys.\  {\bf
36} (1995) 6073, q-alg/9503002.}

\lref\Kontsevichhom{M.~Kontsevich, ``Homological Algebra of Mirror
Symmetry,'' alg-geom/9411018.}

\lref\BondalK{A.I.~Bondal, M.M.~Kapranov, ``Framed triangulated
categories,'' Math. USSR-Sb.  {\bf 70}  (1991) 93.}

\lref\Rastorus{J.~Rasmussen, ``Khovanov's invariant for closed
surfaces,'' math.GT/050252.}

\lref\CSS{J.S.~Carter, M.~Saito, S.~Satoh, ``Ribbon-moves for
2-knots with 1-handles attached and Khovanov-Jacobsson numbers,''
math.GT/0407493.}

\lref\LMnonab{J.~M.~F.~Labastida and M.~Marino, ``NonAbelian
monopoles on four manifolds,'' Nucl.\ Phys.\ B {\bf 448} (1995)
373, hep-th/9504010.}

\lref\PTyurin{V.~Pidstrigach, A.~Tyurin, ``Localisation of the
Donaldson's invariants along Seiberg-Witten classes,''
dg-ga/9507004.}

\lref\Taubeslett{C.~Taubes, ``The Seiberg-Witten and Gromov
invariants,'' Math. Res. Lett. {\bf 2}  (1995) 221.}

\lref\Taubesbook{C.~Taubes, ``Seiberg Witten and Gromov invariants
for symplectic 4-manifolds,''
International Press, Somerville, MA, 2000.}

\lref\Leung{N.~C.~Y.~Leung, ``Topological quantum field theory for
Calabi-Yau threefolds and G(2) manifolds,'' Adv.\ Theor.\ Math.\
Phys.\  {\bf 6} (2003) 575, math.DG/0208124.}

\lref\LeungXW{N.~C.~Y.~Leung, X.~Wang, ``Intersection theory of
coassociative submanifolds in G(2)-manifolds and Seiberg-Witten
invariants,'' math.DG/0401419.}

\lref\ASalur{S.~Akbulut and S.~Salur, ``Calibrated manifolds and
gauge theory,'' math.GT/0402368; ``Associative submanifolds of a
G2 manifold,'' math.GT/0412032.}

\lref\WMoore{ G.~Moore, E.~Witten, ``Integration over the u-plane
in Donaldson theory,'' Adv. Theor. Math. Phys. {\bf 1} (1998) 298,
hep-th/9709193.}

\lref\Dorey{N.~Dorey, ``The BPS Spectra of Two-Dimensional
Supersymmetric Gauge Theories with Twisted Mass Terms,'' JHEP {\bf
9811} (1998) 005, hep-th/9806056.}

\lref\DHT{N.~Dorey, T.J.~Hollowood, D.~Tong, ``The BPS Spectra of
Gauge Theories in Two and Four Dimensions,'' JHEP {\bf 9905}
(1999) 006, hep-th/9902134.}

\lref\SWi{N.~Seiberg and E.~Witten, ``Electric - magnetic duality,
monopole condensation, and confinement in N=2 supersymmetric
Yang-Mills theory,'' Nucl.\ Phys.\ B {\bf 426} (1994) 19
[Erratum-ibid.\ B {\bf 430} (1994) 485], hep-th/9407087.}

\lref\SWii{N.~Seiberg and E.~Witten, ``Monopoles, duality and
chiral symmetry breaking in N=2 supersymmetric QCD,'' Nucl.\
Phys.\ B {\bf 431} (1994) 484, hep-th/9408099.}

\lref\ALabastida{M.~Alvarez, J.M.F.~Labastida, ``Topological
Matter in Four Dimensions,'' Nucl.Phys. {\bf B437} (1995) 356,
hep-th/9404115.}

\lref\WessBagger{J.~Wess, J.~Bagger, ``Supersymmetry and
Supergravity,'' 2nd edition, Princeton University Press, 1992.}

\lref\LLectures{J.~M.~F.~Labastida and C.~Lozano, ``Lectures on
topological quantum field theory,'' hep-th/9709192.}

\lref\Wphases{E.~Witten, ``Phases of N = 2 theories in two
dimensions,'' Nucl.\ Phys.\ B {\bf 403} (1993) 159,
hep-th/9301042.}

\lref\Wgrassmannian{E.~Witten, ``The Verlinde algebra and the
cohomology of the Grassmannian,'' hep-th/9312104.}

\lref\BS{M.~Bershadsky and V.~Sadov, ``Theory of Kahler gravity,''
Int.\ J.\ Mod.\ Phys.\ A {\bf 11}, 4689 (1996), hep-th/9410011.}

\lref\AKMVvertex{M.~Aganagic, A.~Klemm, M.~Marino and C.~Vafa,
``The topological vertex,'' hep-th/0305132.}

\lref\Apol{S.~Gukov, ``Three-dimensional quantum gravity,
Chern-Simons theory, and the  A-polynomial,''
Commun.Math.Phys. {\bf 255} (2005) 577, hep-th/0306165.}

\lref\BCOVa{M.~Bershadsky, S.~Cecotti, H.~Ooguri and C.~Vafa,
``Holomorphic anomalies in topological field theories,'' Nucl.\
Phys.\ B {\bf 405}, 279 (1993).}

\lref\BCOV{M.~Bershadsky, S.~Cecotti, H.~Ooguri and C.~Vafa,
``Kodaira-Spencer theory of gravity and exact results for quantum
string amplitudes,'' Commun.\ Math.\ Phys.\  {\bf 165}, 311
(1994).}

\lref\BJSV{M.~Bershadsky, A.~Johansen, V.~Sadov and C.~Vafa,
``Topological reduction of 4-d SYM to 2-d sigma models,''
Nucl.\ Phys.\ B {\bf 448} (1995) 166, hep-th/9501096.}

\lref\HMStrominger{J.~Harvey, G.~Moore, A.~Strominger,
``Reducing S Duality To T Duality,''
Phys.\ Rev.\  D {\bf 52} (1995) 7161, hep-th/9501022.}

\lref\VafaW{C.~Vafa and E.~Witten, ``A Strong coupling test of S
duality,'' Nucl.\ Phys.\ B {\bf 431} (1994) 3, hep-th/9408074.}

\lref\Manolescu{C.~Manolescu, ``Nilpotent slices, Hilbert schemes,
and the Jones polynomial,'' math.SG/0411015.}

\lref\Manolescusln{C.~Manolescu, ``Link homology theories from
symplectic geometry,'' math.SG/0601629.}

\lref\Bigelow{S.~Bigelow, ``A homological definition of the Jones
polynomial,'' Geom. Topol. Monogr. {\bf 4} (2002) 29,
math.GT/0201221.}

\lref\Beilinson{A.A.~Beilinson, V.G.~Drinfeld, ``Quantization of
Hitchin's fibrations and Langlands' program,'' Math. Phys. Stud.
{\bf 19}, Kluwer Acad. Publ. (1996) 3.}

\lref\DBZias{D.~Ben-Zvi, ``Geometric Langlands as a Quantization
of the Hitchin System,'' lecture at the Princeton workshop on the
Langlands correspondence and physics.}

\lref\CCGLS{D.~Cooper, M.~Culler, H.~Gillet, D.D.~Long, P.B.
Shalen, ``Plane curves associated to character varieties of
3-manifolds,'' Invent. Math. {\bf 118} (1994) 47.}

\lref\Stavros{S.~Garoufalidis, J.~Geronimo, ``Asymptotics of
$q$-difference Equations,'' math.QA/0405331; S.~Garoufalidis,
``Difference and differential equations for the colored Jones
function,'' math.GT/0306229.}

\lref\KWitten{A.~Kapustin, E.~Witten, ``Electric-Magnetic Duality
And The Geometric Langlands Program,'' hep-th/0604151.}

\lref\BTnovel{M.~Blau, G.~Thompson, Nucl.\ Phys.\ B {\bf 492}
(1997) 545; Phys.\ Lett.\ B {\bf 415} (1997) 242.}

\lref\BTbranes{M.~Blau and G.~Thompson, ``Aspects of $N_T \ge 2$
topological gauge theories and D-branes,'' Nucl.\ Phys.\ B {\bf
492} (1997) 545, hep-th/9612143.}

\lref\BTeucl{M.~Blau and G.~Thompson, ``Euclidean SYM theories by
time reduction and special holonomy  manifolds,'' Phys.\ Lett.\ B
{\bf 415} (1997) 242, hep-th/9706225.}

\lref\BTonRW{M.~Blau and G.~Thompson, ``On the relationship
between the Rozansky-Witten and the three-dimensional
Seiberg-Witten invariants,'' Adv.\ Theor.\ Math.\ Phys.\  {\bf 5}
(2002) 483, hep-th/0006244.}

\lref\GMext{B.~Geyer, D.~Mülsch, ``$N_T=4$ equivariant extension
of the 3D topological model of Blau and Thompson,'' Nucl.Phys.
{\bf B616} (2001) 476, hep-th/0108042.}

\lref\MMsympl{M.~Marino and G.~W.~Moore, ``Donaldson invariants
for non-simply connected manifolds,'' Commun.\ Math.\ Phys.\  {\bf
203} (1999) 249, hep-th/9804104.}

\lref\MMthreed{M.~Marino and G.~W.~Moore, ``3-manifold topology
and the Donaldson-Witten partition function,'' Nucl.\ Phys.\ B
{\bf 547} (1999) 569, hep-th/9811214.}

\lref\MMcoulomb{M.~Marino and G.~W.~Moore,
``Integrating over the Coulomb branch in N = 2 gauge theory,''
Nucl.\ Phys.\ Proc.\ Suppl.\  {\bf 68} (1998) 336, hep-th/9712062.}

\lref\MMrank{M.~Marino and G.~W.~Moore,
``The Donaldson-Witten function for gauge groups of rank larger than one,''
Commun.\ Math.\ Phys.\  {\bf 199} (1998) 25, hep-th/9802185.}

\lref\MMPeradze{M.~Marino, G.~W.~Moore and G.~Peradze,
``Superconformal invariance and the geography of four-manifolds,''
Commun.\ Math.\ Phys.\  {\bf 205} (1999) 691, hep-th/9812055.}

\lref\Issues{A.~Losev, N.~Nekrasov and S.~L.~Shatashvili,
``Issues in topological gauge theory,''
Nucl.\ Phys.\ B {\bf 534} (1998) 549, hep-th/9711108.}

\lref\RW{L.~Rozansky and E.~Witten, ``Hyper-Kaehler geometry and
invariants of three-manifolds,'' Selecta Math.\  {\bf 3} (1997)
401, hep-th/9612216.}

\lref\Seidel{P.~Seidel, ``Lagrangian two-spheres can be symplectically knotted,''
J. Diff. Geom. {\bf 52} (1999) 145, math.DG/9803083.}

\lref\SThomas{P.~Seidel, R.~Thomas, ``Braid group actions on derived categories
of coherent sheaves,'' Duke Math. J. {\bf 108} (2001) 37, math.AG/0001043.}

\lref\DeligneBraid{P.~Deligne, ``Action du groupe des tresses sur une categorie,''
Invent. Math. {\bf 128} (1997) 159.}

\lref\Simpson{C.~Simpson, ``Nonabelian Hodge theory,'' Proc. I.C.M., Kyoto 1990,
Springer-Verlag, 1991, pp. 198-230.}

\lref\IwasakiU{K.~Iwasaki, T.~Uehara, ``Periodic Solutions
to Painleve VI and Dynamical System on Cubic Surface,'' math.AG/0512583.}

\lref\IISaito{M.~Inaba, K.~Iwasaki and M.-H.~Saito,
``Dynamics of the sixth Painleve equation,''
Proceedings of Conference Internationale Theories Asymptotiques
et Equations de Painleve, Seminaires et Congres, Soc. Math. France, math.AG/0501007.}

\lref\Iwasaki{K.~Iwasaki, ``An area-preserving action of the modular group
on cubic surfaces and the Painleve VI equation,'' Comm. Math. Phys. {\bf 242} (2003) 185.}

\lref\Iwasakid{K.~Iwasaki, ``A modular group action on cubic surfaces and the monodromy
of the Painleve VI equation,'' Proc. Japan Acad. Ser. A Math. Sci. {\bf 78} (2002) 131.}

\lref\Sakai{H.~Sakai, ``Rational surfaces associated with affine root systems
and geometry of the Painleve equations,'' Comm. Math. Phys. {\bf 220}  (2001) 165.}

\lref\HitSchles{N.~Hitchin, ``Geometrical aspects of Schlesinger's equation,''
J. Geom. Phys. {\bf 23}  (1997) 287.}

\lref\Lin{X.S.~Lin, ``A knot invariant via representation spaces,''
J. Diff. Geom. {\bf 35} (1992) 337.}

\lref\Kroll{J.~Kroll,  ``\"Aquivariante Signatur und
$SU(2)$- Darstellungsr\"aume von Knotengruppen,''
Diplomarbeit, Universität-Gesamthochschule Siegen, 1996.}

\lref\Herald{C.~Herald, ``Existence of irreducible representations
of knot complements with nonconstant equivariant signature,''
Math. Ann. {\bf 309} (1997) 21.}

\lref\Heraldi{C.~Herald, ``Flat connections, the Alexander invariant,
and Casson's invariant,'' Comm. Anal. Geom. {\bf 5}  (1997) 93.}

\lref\CollinSteer{O.~Collin, B.~Steer, ``Instanton Floer homology for knots
via 3-orbifolds,'' J. Diff. Geom. {\bf 51} (1999) 149.}

\lref\Collinii{O.~Collin, ``Floer Homology for Knots and $SU(2)$-Representations
for Knot Complements and Cyclic Branched Covers,'' Canad. J. Math. {\bf 52} (2000) 293.}

\lref\Collin{O.~Collin, ``Floer Homology for Orbifolds and
Gauge Theory Knot Invariants,'' {\it Knots 96,}
World Scientific Publishing Co., Singapore (1997) 201.}

\lref\Klassen{E.~Klassen, ``Representations of knot groups in $SU(2)$,''
Trans. Amer. Math. Soc. {\bf 326} (1991) 795.}

\lref\BHKK{H.~Boden, C.~Herald, P.~Kirk, E.~Klassen,
``Gauge Theoretic Invariants of Dehn Surgeries on Knots,''
Geom. Topol. {\bf 5} (2001) 143, math.GT/9908020.}

\lref\LicassonL{W.~Li, ``Casson-Lin's invariant and floer homology,''
J. Knot Theory and its Ramification, {\bf 6} (1997) 851.}

\lref\WLi{W.~Li, ``Knot and link invariants and moduli space
of parabolic bundles,'' Commun. Contemp. Math. {\bf 3} (2001) 501.}

\lref\DonaldsonMT{S.K.~Donaldson,
``Topological field theories and formulae of Casson and Meng-Taubes,''
Geom. Topol. Monogr.{\bf 2} (1999) 87.}

\lref\Bradlow{S.~Bradlow, ``Special metrics and stability for holomorphic bundles
with global sections,'' J. Diff. Geom. {\bf 33} (1991) 169.}

\lref\OSadjunction{P.~Ozsvath, Z.~Szabo, ``Higher type adjunction inequalities
in Seiberg-Witten theory,'' J. Diff. Geom. {\bf 55} (2000) 385.}

\lref\MrowkaOY{T.~Mrowka, P.~Ozsvath, B.~Yu, ``Seiberg-Witten monopoles on Seifert
fibered spaces,'' Comm. Anal. Geom. {\bf 5} (1997) 685.}

\lref\MorganST{J.~Morgan, Z.~Szabo, C.~Taubes, ``A product formula for the Seiberg-Witten
invariants and the generalized Thom conjecture,''  J. Diff. Geom. {\bf 44} (1996) 706.}

\lref\MarkSW{T.~Mark, ``Torsion, TQFT, and Seiberg-Witten invariants of 3-manifolds,''
Geom.Topol. {\bf 6} (2002) 27.}

\lref\MunozWang{V.~Munoz, B.-L.~Wang, ``Seiberg-Witten-Floer homology of a surface
times a circle for non-torsion ${\rm spin}\sp {\Bbb C}$ structures,''
Math. Nachr. {\bf 278}  (2005) 844.}

\lref\NielsenOlesen{H.~B.~Nielsen and P.~Olesen,
``Vortex-Line Models For Dual Strings,''  Nucl.\ Phys.\ B {\bf 61} (1973) 45}

\lref\KMi{P.~Kronheimer, T.~Mrowka, ``Gauge theory for embedded surfaces. I,''
Topology {\bf 32} (1993) 773.}

\lref\KMii{P.~Kronheimer, T.~Mrowka, ``Gauge theory for embedded surfaces. II,''
Topology {\bf 34} (1995) 37.}

\lref\KMstructure{P.~Kronheimer, T.~Mrowka, ``Embedded surfaces and the structure
of Donaldson's polynomial invariants,'' J. Diff. Geom. {\bf 41} (1995) 573.}

\lref\WittenAbelian{ E.~Witten, ``On S duality in Abelian gauge theory,''
Selecta Math.\  {\bf 1} (1995) 383, hep-th/9505186.}

\lref\VerlindeAbelian{E.~Verlinde, ``Global aspects of electric - magnetic duality,''
Nucl.\ Phys.\ B {\bf 455} (1995) 211, hep-th/9506011.}

\lref\Gottsche{L.~G\"{o}ttsche, ``The Betti numbers of the Hilbert scheme of points
on a smooth projective surface,''  Math. Ann. {\bf 286} (1990) 193.}

\lref\GWramified{S.~Gukov and E.~Witten,
``Gauge theory, ramification, and the geometric Langlands program,'' hep-th/0612073.}

\lref\Hitchin{N.~Hitchin, ``The Self-Duality Equations On A Riemann Surface,''
Proc. London Math. Soc. (3) {\bf 55} (1987) 59.}

\lref\Maldacena{J.~M.~Maldacena,
``The large N limit of superconformal field theories and supergravity,''
Adv.\ Theor.\ Math.\ Phys.\  {\bf 2} (1998) 231, hep-th/9711200.}

\lref\inprogress{work in progress}

\lref\Lim{Y.~Lim, ``The equivalence of Seiberg-Witten and Casson invariants
for homology 3-spheres,'' Math. Res. Lett. {\bf 6} (1999), 631.}

\lref\CLMiller{S.~Cappell, R.~Lee, E.~Miller,
``Surgery formulae for analytical invariants of manifolds,''
Contemp. Math. {\bf 279}, Amer. Math. Soc., Providence, RI, 2001.}

\lref\Nicolaescu{L.~Nicolaescu, ``Seiberg-Witten invariants of rational
homology spheres,'' math.GT/0103020.}

\lref\CLMillerCasson{S.~Cappell, R.~Lee, E.~Miller, ``Equivariant Casson
invariant,'' preprint.}

\lref\GWalcher{S.~Gukov, J.~Walcher, ``Matrix Factorizations and Kauffman
Homology,'' hep-th/0512298.}

\lref\GIKV{S.~Gukov, A.~Iqbal, C.~Kozcaz, C.~Vafa,
``Link Homologies and the Refined Topological Vertex,'' arXiv:0705.1368.}

\lref\BCurtis{H.~Boden, C.~Curtis, ``The $SL(2,C)$ Casson invariant for
Seifert fibered homology spheres and surgeries on twist knots,'' math.GT/0602023.}

\def\boxit#1{\vbox{\hrule\hbox{\vrule\kern8pt
\vbox{\hbox{\kern8pt}\hbox{\vbox{#1}}\hbox{\kern8pt}}
\kern8pt\vrule}\hrule}}
\def\mathboxit#1{\vbox{\hrule\hbox{\vrule\kern8pt\vbox{\kern8pt
\hbox{$\displaystyle #1$}\kern8pt}\kern8pt\vrule}\hrule}}


\let\includefigures=\iftrue
\newfam\black
\includefigures
\input epsf
\def\figin{\epsfcheck\figin}\def\figins{\epsfcheck\figins}
\def\epsfcheck{\ifx\epsfbox\UnDeFiNeD
\message{(NO epsf.tex, FIGURES WILL BE IGNORED)}
\gdef\figin##1{\vskip2in}\gdef\figins##1{\hskip.5in}
\else\message{(FIGURES WILL BE INCLUDED)}%
\gdef\figin##1{##1}\gdef\figins##1{##1}\fi}
\def\DefWarn#1{}
\def\figinsert{\goodbreak\midinsert}
\def\ifig#1#2#3{\DefWarn#1\xdef#1{fig.~\the\figno}
\writedef{#1\leftbracket fig.\noexpand~\the\figno}%
\figinsert\figin{\centerline{#3}}\medskip\centerline{\vbox{\baselineskip12pt
\advance\hsize by -1truein\noindent\footnotefont{\bf
Fig.~\the\figno:} #2}}
\bigskip\endinsert\global\advance\figno by1}
\else
\def\ifig#1#2#3{\xdef#1{fig.~\the\figno}
\writedef{#1\leftbracket fig.\noexpand~\the\figno}%
\global\advance\figno by1} \fi

\newdimen\tableauside\tableauside=1.0ex
\newdimen\tableaurule\tableaurule=0.4pt
\newdimen\tableaustep
\def\phantomhrule#1{\hbox{\vbox to0pt{\hrule height\tableaurule width#1\vss}}}
\def\phantomvrule#1{\vbox{\hbox to0pt{\vrule width\tableaurule height#1\hss}}}
\def\sqr{\vbox{%
  \phantomhrule\tableaustep
  \hbox{\phantomvrule\tableaustep\kern\tableaustep\phantomvrule\tableaustep}%
  \hbox{\vbox{\phantomhrule\tableauside}\kern-\tableaurule}}}
\def\squares#1{\hbox{\count0=#1\noindent\loop\sqr
  \advance\count0 by-1 \ifnum\count0>0\repeat}}
\def\tableau#1{\vcenter{\offinterlineskip
  \tableaustep=\tableauside\advance\tableaustep by-\tableaurule
  \kern\normallineskip\hbox
    {\kern\normallineskip\vbox
      {\gettableau#1 0 }%
     \kern\normallineskip\kern\tableaurule}%
  \kern\normallineskip\kern\tableaurule}}
\def\gettableau#1 {\ifnum#1=0\let\next=\null\else
  \squares{#1}\let\next=\gettableau\fi\next}

\tableauside=1.0ex \tableaurule=0.4pt

\font\cmss=cmss10 \font\cmsss=cmss10 at 7pt

\def\IB{\relax\hbox{$\inbar\kern-.3em{\rm B}$}}
\def\IC{\relax\hbox{$\inbar\kern-.3em{\rm C}$}}
\def\IQ{\relax\hbox{$\inbar\kern-.3em{\rm Q}$}}
\def\ID{\relax\hbox{$\inbar\kern-.3em{\rm D}$}}
\def\IE{\relax\hbox{$\inbar\kern-.3em{\rm E}$}}
\def\IF{\relax\hbox{$\inbar\kern-.3em{\rm F}$}}
\def\IG{\relax\hbox{$\inbar\kern-.3em{\rm G}$}}
\def\IGa{\relax\hbox{${\rm I}\kern-.18em\Gamma$}}
\def\IH{\relax{\rm I\kern-.18em H}}
\def\IK{\relax{\rm I\kern-.18em K}}
\def\IL{\relax{\rm I\kern-.18em L}}
\def\IP{\relax{\rm I\kern-.18em P}}
\def\IR{\relax{\rm I\kern-.18em R}}
\def\Z{\relax\ifmmode\mathchoice
{\hbox{\cmss Z\kern-.4em Z}}{\hbox{\cmss Z\kern-.4em Z}}
{\lower.9pt\hbox{\cmsss Z\kern-.4em Z}} {\lower1.2pt\hbox{\cmsss
Z\kern-.4em Z}}\else{\cmss Z\kern-.4em Z}\fi}

\def\II{\relax{\rm I\kern-.18em I}}

\def\S{{\bf S}}
\def\B{{\bf B}}
\def\R{{\bf R}}
\def\C{{\bf C}}

\def\cp{{\bf CP}}

\def\CA {{\cal A}}
\def\CB {{\cal B}}

\def\CE {{\cal E}}
\def\CF {{\cal F}}

\def\CH {{\cal H}}

\def\CL {{\cal L}}
\def\CM {{\cal M}}
\def\CN {{\cal N}}
\def\CO {{\cal O}}


\def\p{\partial}

\def\D{{\slash\!\!\!\! D}}

\def\tilde{\widetilde}

\def\bar{\overline}


\def\Tr{{\rm Tr}}

\def\p{\partial}

\def\lieg{{\bf g}}

\def\lieg{\frak g}

\def\CCG{G_{\scriptscriptstyle{{\bf C}}}}

\def\inbar{\,\vrule height1.5ex width.4pt depth0pt}
\def\r{{\rm Re}}

\def\a{\alpha}
\def\b{\beta}
\def\g{\gamma}

\def\la{\lambda}

\def\bar{\overline}

\def\det{{\rm det}}
\def\tr{{\rm tr}}

\def\Tr{{\rm Tr}}

\def\IH{{\bf H}}

\def\leadsto{\rightsquigarrow}

\def\example#1{\bgroup\narrower\footnotefont\baselineskip\footskip\bigbreak
\hrule\medskip\nobreak\noindent {\bf Example}. {\it
#1\/}\par\nobreak}
\def\endexample{\medskip\nobreak\hrule\bigbreak\egroup}

\def\TT{{\Bbb{T}}}

\def\Weyl{{\cal W}}


\noindent
\Title{\vbox{\baselineskip12pt\hbox{ITEP-TH-27/07}
\hbox{LANDAU-07-01}
}} {\vbox{
\centerline{Surface Operators and Knot Homologies}}}
\bigskip
\smallskip
\centerline{Sergei Gukov}
\smallskip
\centerline{\it{Department of Physics, University of California}}
\centerline{\it{Santa Barbara, CA 93106}}
\medskip
\bigskip\bigskip
\noindent
Topological gauge theories in four dimensions which admit surface operators
provide a natural framework for realizing homological knot invariants.
Every such theory leads to an action of the braid group on branes on
the corresponding moduli space.
This action plays a key role in the construction of homological knot invariants.
We illustrate the general construction with examples based
on surface operators in $\CN=2$ and $\CN=4$ twisted gauge theories
which lead to a categorification of
the Alexander polynomial,
the equivariant knot signature, and certain analogs of the Casson invariant.

This paper is based on a lecture delivered
at the International Congress on Mathematical Physics 2006,
Rio de Janeiro, and at the RTN Workshop 2006, Napoli.
\vskip .5cm
\noindent\Date{June, 2007}




\newsec{Introduction}

Topological field theory is a natural framework for ``categorification'',
an informal procedure that turns integers into vector spaces
(abelian groups), vector spaces into abelian or triangulated categories,
operators into functors between these categories \CFrenkel.
The number becomes the dimension of the vector space,
while the vector space becomes the Grothendieck group
of the category (tensored with a field). This procedure can
be illustrated by the following diagram \HKhovanov:
\medskip
\eqn\categorificationchart{
{\lower3.0pt \hbox{\epsfxsize3.5in\epsfbox{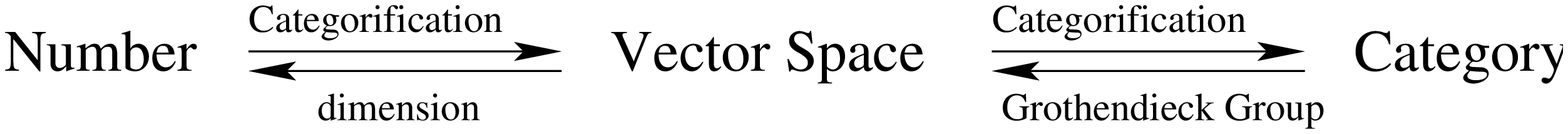}}} }
Recently, this idea led to a number of remarkable developments in
various branches of mathematics, notably in low-dimensional topology,
where many polynomial knot invariants were lifted to homological invariants.

Although the list of homological knot invariants is constantly growing,
most of the existing knot homologies fit into the ``A-series''
of homological knot invariants
associated with the fundamental representation of $sl(N)$ (or $gl(N)$).
Each such theory is a doubly graded knot homology whose graded Euler
characteristic with respect to one of the gradings gives the corresponding
knot invariant,
\eqn\pfromh{ P(q) = \sum_{i,j} (-1)^i q^j \dim \CH_{i,j} }
For example, the Jones polynmial can be obtained in this way
as the graded Euler characteristic of the Khovanov homology \Khovanov.
Similarly, the so-called knot Floer homology \refs{\OShfk,\Rasmussen}
provides a categorification of the Alexander polynomial $\Delta (q)$.
At first, these as well as other homological knot invariants listed in
the table below appear to have very different character.
Thus, as the name suggests, knot Floer homology is defined as a symplectic
Floer homology of two Lagrangian submanifolds in a certain configuration space,
while the other theories are defined combinatorially.
In addition, the definition of the knot Floer homology admits a generalization
to knots in arbitrary 3-manifolds, whereas at present the definition
of the other knot homologies (with $N>0$) is known only for knots in $\R^3$.


\vskip 0.8cm \vbox{ \centerline{\vbox{
\hbox{\vbox{\offinterlineskip
\def\tablespace{height7pt&\omit&&\omit&&\omit&\cr}
\def\tablerule{\tablespace\noalign{\hrule}\tablespace}

\hrule\halign{&\vrule#&\strut\hskip0.2cm\hfill
#\hfill\hskip0.2cm\cr
\tablespace & $\lieg$ && Knot Polynomial && Categorification &\cr
\tablerule & $gl(1|1)$ && $\Delta (q)$ && knot Floer homology $HFK
(K)$ &\cr \tablerule & ``$sl(1)$'' && ---  && Lee's deformed
theory $H'(K)$ &\cr \tablerule & $sl(2)$ && Jones && Khovanov
homology $H^{{\rm Kh}} (K)$ &\cr \tablerule & $sl(N)$ && $P_N (q)$
&& $sl(N)$ homology $HKR^N (K)$ &\cr
\tablespace}\hrule}}}}
\centerline{ \hbox{{\bf Table 1:}{\it ~~ A general picture of knot
polynomials and knot homologies.}}}
} \vskip 0.5cm

The $sl(N)$ knot homology \refs{\Khovanov,\Khovanoviii,\RKhovanov}
--- whose Euler characteristic is the quantum $sl(N)$ invariant $P_N (q)$ ---
has a physical interpretation as the space
of BPS states, $\CH_{BPS}$, in string theory \GSV.
In order to remind the physical setup of \GSV, let us recall that
polynomial knot invariants, such as $P_N (q)$, can be related to
open topological srting amplitudes (``open Gromov-Witten invariants'')
by first embedding Chern-Simons gauge theory in topological string theory \Wittencsstring,
and then using the so-called large $N$ duality \refs{\GViii,\OV,\LMV},
a close cousin of the celebrated AdS/CFT duality \Maldacena.
Moreover, open topological string amplitudes and, hence, the corresponding
knot invariants can be reformulated in terms of new integer invariants
which capture the spectrum of BPS states in the string Hilbert space, $\CH_{BPS}$.
The BPS states in question are membranes ending on Lagrangian five-branes
in M-theory on a non-compact Calabi-Yau space
$\eusm X = \CO_{{\bf P^1}} (-1) \oplus \CO_{{\bf P^1}} (-1)$.
Specifically, the five-branes have world-volume ${\bf R}^{2,1} \times L_K$
where $L_K \subset \eusm X$ is a Lagrangian submanifold (which depends on knot $K$)
and ${\bf R}^{2,1} \subset {\bf R}^{4,1}$.

\ifig\curv{A membrane ending on a Lagrangian five-brane.}{\epsfxsize2.8in\epsfbox{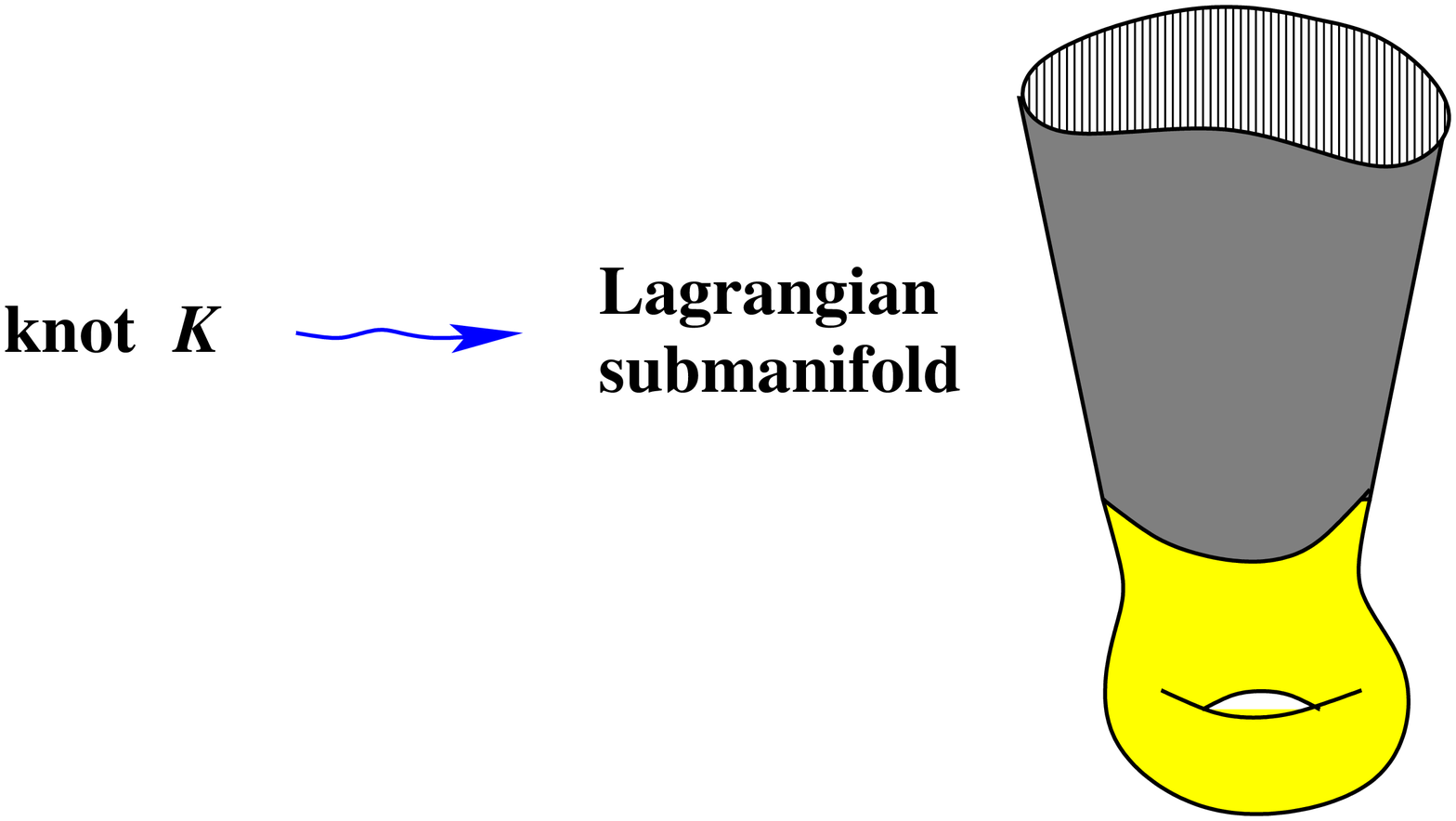}}

Surprisingly, the physical interpretation of the $sl(N)$ knot homology
naturally leads to a triply-graded (rather than doubly-graded)
knot homology \GSV\ (see also \refs{\GWalcher,\GIKV}).
Indeed, the Hilbert space of BPS states, $\CH_{BPS}$, is graded by three
quantum numbers, which are easy to see in the physical setup described
in the previous paragraph. The world-volume of the five-brane breaks the
$SO(4) \cong SU(2) \times SU(2)$ rotation symmetry in five dimensions
down to a subgroup $U(1)_L \times U(1)_R$, where $U(1)_L$ (resp. $U(1)_R$)
is a rotation symmetry in the dimensions parallel (resp. transverse) to the five-brane.
Therefore, BPS states in the effective $\CN=2$ theory on the five-brane
are labeled by three quantum numbers $j_L$, $j_R$, and $Q$,
where $Q \in H_2 (\eusm X, L_K) \cong \Z$ is the relative homology class
represented by the membrane world-volume.
In other words, apart from the $\Z_2$-grading by the fermion number,
the Hilbert space of BPS states $\CH_{BPS}$ is triply-graded.
The properties of this triply-graded theory were studied in \DGR;
it turns out that this theory unifies all the doubly-graded
knot homologies listed in Table 1, including the knot Floer homology.
A mathematical definition of the triply-graded knot homology
which appears to have many of the expected properties was constructed in \RKhovanovii.

Apart from realization in (topological) string theory, the homological
knot invariants are expected to have a physical realization
also in topological gauge theory, roughly as polynomial knot invariants
have a physical realization in three-dimensional gauge theory
(namely, in the Chern-Simons theory \WittenJones)
as well as in the topological string theory \refs{\Wittencsstring,\GViii,\OV}.
Although these two realizations are not unrelated, different properties
of knot polynomials are easier to see in one description or the other.
For example, the dependence on the rank $N$ is manifest in
the string theory description, while the skein operations and transformations
under surgeries are easier to see in the Chern-Simons gauge theory.

Similarly, as we explained above, string theory realization is very useful
for understanding relation between knot homologies of different rank.
On the other hand, the formal properties of knot homologies which
are hard to see in string theory
(which, however, would be very natural in topological field theory)
have to do with the fact that, in most cases, knot homologies can be
extended to a functor $\CF$ from the category of 3-manifolds with links
and cobordisms to the category of graded vector spaces and homomorphisms
\eqn\ykfunctor{ \CF (Y;K) = \CH_{Y;K} }
\eqn\ykmapfunctor{ \CF (X;D) ~:~~ \CH_{Y;K} \to \CH_{Y';K'} }
Moreover, on manifolds with corners, it is expected that $\CF$
extends to a 2-functor from the 2-category of oriented and
decorated 4-manifolds with corners to the 2-category of
triangulated categories \refs{\Khovanovtangles,\Jacobsson,\DBNtangles}.
In particular, it should associate:

$\bullet$ a triangulated category $\CF (\Sigma)$ to a closed
oriented 2-manifold $\Sigma$;

$\bullet$ an exact functor $\CF (Y)$ to a 3-dimensional oriented
cobordism $Y$;

$\bullet$ a natural transformation $\CF (X)$ to a 4-dimensional
oriented cobordism $X$.

\noindent
As we explain below, these are precisely the formal properties
of a four-dimensional topological field theory with boundaries and corners.
Moreover, links and link cobordisms can be incorporated by introducing
``surface operators'' in the topological gauge theory.

In section 2, we discuss the general aspects of topological gauge
theories which admit surface operators.
Of particular importance is the fact that every topological gauge theory
which admits surface operators gives rise to an action of the braid group on D-branes.
Then, in sections 3 and 4 we illustrate how these general structures are realized
in simple examples of $\CN=2$ and $\CN=4$ twisted gauge theories.
Specifically, in section 3 we study surface operators and the corresponding
knot homologies in the Donaldson-Witten theory and in the Seiberg-Witten theory,
both of which are obtained by twisting $\CN=2$ supersymmetric gauge theory.
In section 4, we explain that a particular twist of the $\CN=4$ super-Yang-Mills
theory --- studied recently in connection with the geometric Langlands
program \refs{\KWitten,\GWramified} --- with a simple type of surface
operators provides a physical framework for categorification of
the $\CCG$ Casson invariant.


\newsec{Gauge Theory and Categorification}

Let us start by describing the general properties
of the topological quantum field theory (TQFT) with boundaries,
corners, and surface operators.
To a closed 4-manifold $X$, a four-dimensional TQFT
associates a number, $Z(X)$, the partition function of the
topological theory on $X$. Similarly, to a closed 3-manifold $Y$,
it associates a vector space, $\CH_Y$, the Hilbert space obtained
by quantization of the theory on $X = \R \times Y$. Finally, to a
closed surface $\Sigma$ it associates a triangulated category,
$\CF(\Sigma)$, which can be understood as the category of D-branes
in the topological sigma-model obtained via the dimensional
reduction of gauge theory on $\Sigma$. The objects of the category
$\CF(\Sigma)$ describe BRST-invariant boundary conditions in
the four-dimensional TQFT on 4-manifolds with corners
(locally, such manifolds look like $X = \R \times \R_+ \times \Sigma$).
Summarizing,
\vskip 1em \centerline{\vbox{\halign{\quad # & \quad # & \quad
#\cr
gauge theory on $X$ & $\leadsto$ & number $Z(X)$ \cr
gauge
theory on $\R \times Y$ & $\leadsto$ & vector space $\CH_Y$ \cr
gauge theory on $\R^2 \times \Sigma$ & $\leadsto$ & category $\CF
(\Sigma)$ \cr
 }}}
\vskip 0.3em
\noindent
where we assume that $X$, $Y$, and $\Sigma$ are closed.
Depending on whether the topological
reduction of the four-dimensional gauge theory on $\R^2 \times \Sigma$
gives a topological A-model or B-model, the category $\CF(\Sigma)$
is either the derived Fukaya category\foot{Notice, according
to the Homological Mirror Symmetry conjecture, this category
is equivalent to the derived category of the mirror B-model \Kontsevichhom.
In particular, the category ${\rm\bf Fuk} (\CM)$, suitably defined,
must be a triangulated category \BondalK.}, ${\rm\bf Fuk} (\CM)$, or the derived
category of coherent sheaves, $D^b (\CM):=D^b {\rm\bf Coh} (\CM)$,
\vskip 1em \centerline{\vbox{\halign{\quad # & \quad #\cr
topological A-model: & $\CF (\Sigma) = {\rm\bf Fuk} (\CM)$ \cr
topological B-model: & $\CF (\Sigma) = D^b (\CM)$ \cr
 }}}
\vskip 0.3em \noindent where $\CM$ is the moduli space of
classical solutions in gauge theory on $\R^2 \times \Sigma$,
invariant under translations along $\R^2$.
Different topological gauge theories lead to different functors $\CF$.
For example, in the context of Donaldson-Witten theory \WittenDonaldson,
Fukaya suggested \Fukaya\ that the category associated
to a closed surface $\Sigma$ should be $A_{\infty}$-category
of Lagrangian submanifolds in the moduli space of flat
$G$-connections on $\Sigma$. This is precisely what
one finds from the topological reduction \BJSV\
of the twisted $\CN=2$ gauge theory on $\R^2 \times \Sigma$,
in agreement with the general principle discussed here.

\medskip\noindent{{\it The Atiyah-Floer conjecture and its variants}}\medskip

It is easy to see that $\CF$ defined by the topological gauge
theory has all the expected properties of a 2-functor.
In particular, to a 3-manifold $Y$ with boundary $\p Y = \Sigma$ it associates a
``D-brane'', that is an object in the category $\CF (\Sigma)$.

\ifig\munion{
$a)$ A 3-manifold $Y$ can be obtained as a connected sum of 3-manifolds
$Y_1$ and $Y_2$, joined along their common boundary $\Sigma$.
$b)$ In four-dimensional gauge theory, the space $\R \times Y$
is obtained by gluing two 4-manifolds with corners.} {\epsfxsize4.5in\epsfbox{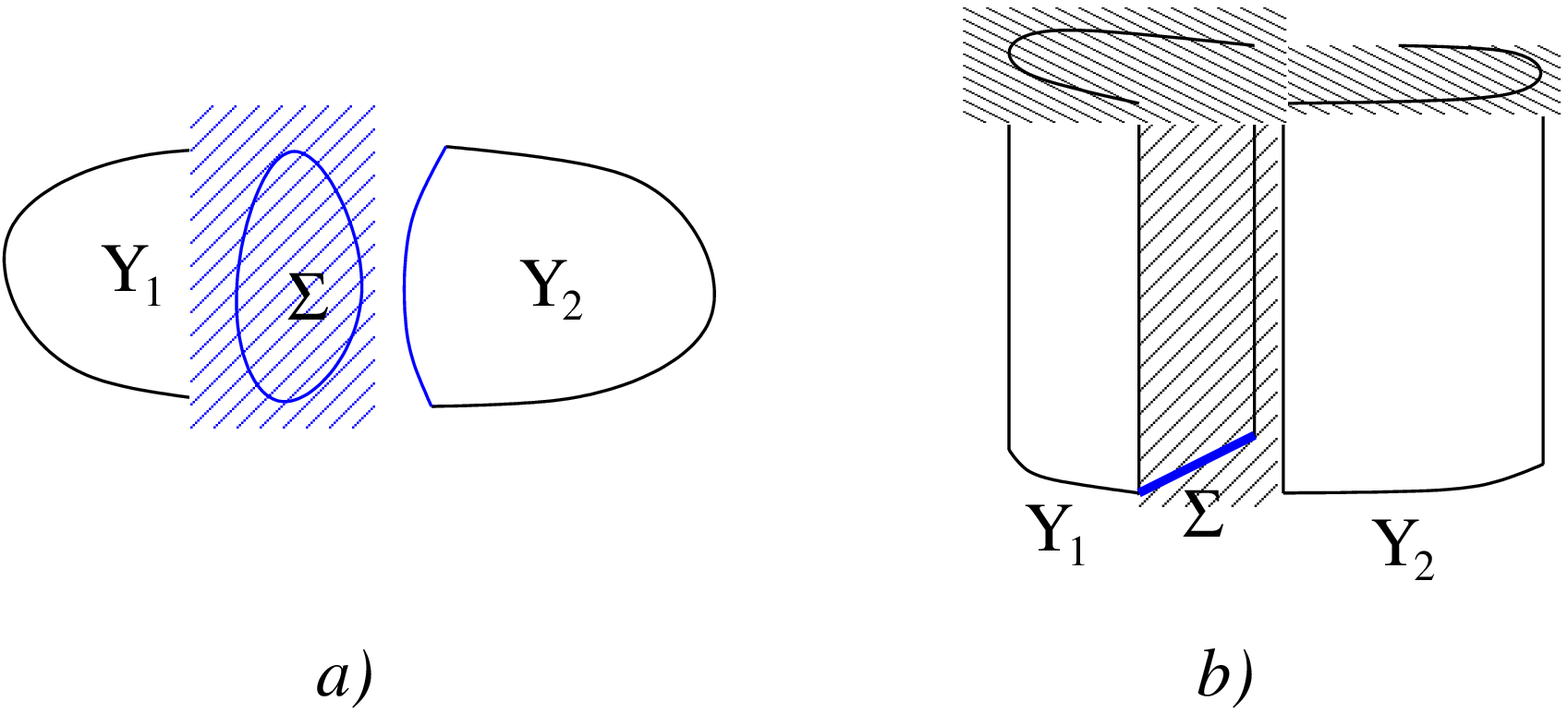}}

The interpretation of 3-manifolds with boundary as D-branes
can be used to reproduce the Atiyah-Floer conjecture,
which states \AtiyahF:
\eqn\AFconj{ HF^{{\rm inst}}_* (Y) \cong HF^{{\rm symp}}_* (\CM;
\CL_{1}, \CL_{2}) }
Here $\CM = \CM_{flat}^{G}$ is the moduli space of flat connections
on $\Sigma$, while $\CL_{1}$ and $\CL_{2}$ are Lagrangian
submanifolds in $\CM$ associated with the Heegard splitting of $Y$,
\eqn\heegard{ Y = Y_1 \cup_{\Sigma} Y_2 }
such that the points of $\CL_i \subset \CM$, $i=1,2$,
correspond to flat connections on $\Sigma$ which can be extended to $Y_i$.

Similarly, in the B-model, $Y_1$ and $Y_2$ define the corresponding B-branes,
which are objects in the derived category of coherent sheaves on $\CM$.
In both cases, the vector space $\CH_Y$ associated with the compact
3-manifold $Y$ is the space of ``$1-2$ strings'':
\eqn\abstrings{ \CH_Y 
= \cases{ HF^{{\rm symp}}_* (\CM; \CL_{1}, \CL_{2})
& A-model \cr {\rm Ext}^* (\CF_{Y_1},\CF_{Y_2}) & B-model} }
In the Donaldson-Witten theory, this leads to the Atiyah-Floer conjecture \AFconj.

\medskip\noindent{{\it ``Decategorification''}}\medskip

The operation represented by the arrow going to the left
in \categorificationchart\ --- ``decategorification'' --- also
has a natural interpretation in gauge theory. It corresponds
to the dimensional reduction, or compactification on a circle.
Indeed, the partition function in gauge theory on $X = \S^1 \times Y$
is the trace (the index) over the Hilbert space $\CH_Y$:
\eqn\circlea{ Z_{\S^1 \times Y} = \chi (\CH_Y) }
Similarly, the vector space associated with $Y = \S^1 \times \Sigma$
is the Grothendieck group of the category $\CF (\Sigma)$
\eqn\circleb{ \CH_{\S^1 \times \Sigma} = K (\CF (\Sigma)) }
In the case of A-model and B-model, respectively, we find
\eqn\grgroup{ K(\CF (\Sigma)) = \cases{H^{d} (\CM) & for $\CF (\Sigma) = {\rm\bf
Fuk} (\CM)$ \cr H^* (\CM) & for $\CF (\Sigma) = D^b (\CM)$ }}
where $d = \half \dim (\CM)$.


\subsec{Incorporating Surface Operators}

In a three-dimensional TQFT, knots and links can be incorporated
by inrtoducing topological loop observables. The familiar example
is the Wilson loop observable in Chern-Simons theory,
\eqn\wloop{ W_R (K) = \Tr_R \Big( P \exp \oint_{K} A \Big) }
Recall, that canonical quantization of the Chern-Simons theory on
$\Sigma \times \R$ associates a vector space $\CH_{\Sigma}$ ---
the ``physical Hilbert space'' --- to a Riemann surface $\Sigma$ \WittenJones.
In presence of Wilson lines, quantization gives a Hilbert space
$\CH_{\Sigma;p_i,R_i}$ canonically associated to a Riemann
surface $\Sigma$ together with marked points $p_i$
(points where Wilson lines meet $\Sigma$) decorated by representations $R_i$.
For example, to $n$ marked points on the plane colored by
the fundamental representation it associates $\eurm V^{\otimes n}$,
where $\eurm V$ is a $N$-dimensional irreducible representation of the
quantum group $U_q (sl(N))$.

\ifig\linesfig{$a)$ line operators in a three-dimensional TQFT on $\Sigma \times I$ and
$b)$ topological ``surface operators'' in four-dimensional gauge theory
on $\Sigma \times I \times \R$.} {\epsfxsize4.7in\epsfbox{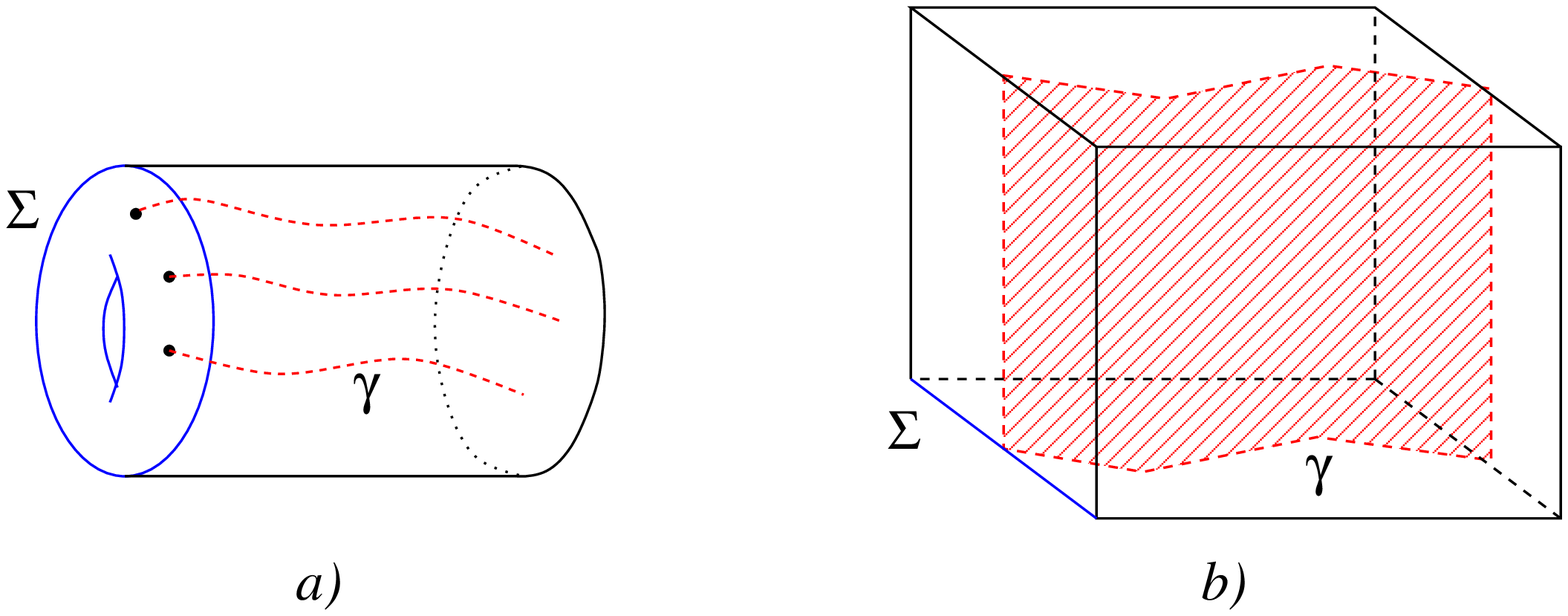}}

We wish to lift this to a four-dimensional gauge theory by including
the ``time'' direction, so that the space-time becomes $X = Y \times \R$,
where the knot $K$ is represented by a topological defect
(which was called a ``surface operator'' in \GWramified)
localized on the surface $D = K \times \R$.
In the Feynman path integral, a surface operator is defined by requiring
the gauge field $A$ (and perhaps other fields as well) to have a prescribed singularity.
For example, the simplest type of singularity studied in \GWramified\
creates a holonomy of the gauge field on a small loop around $D$,
\eqn\vhol{ V = {\rm Hol} (A) }
Quantization of the four-dimensional topological theory on
a 4-manifold $X = Y \times \R$ with a surface operator on $D = K \times \R$
gives rise to a functor that associates to this data
(namely, a 3-manifold $Y$, a knot $K$, and parameters of the surface operator)
a vector space, the space of quantum ground states,
\eqn\ykfunctor{ \CF (Y;K) = \CH_{Y;K,{\rm parameters}} }
Moreover, we will be interested in surface operators
which preserve topological invariance for more general
4-manifolds $X$ and embedded surfaces $D \subset X$.
For example, if the four-dimensional topological gauge theory
is obtained by a topological twist of a supersymmetric gauge theory,
it is natural to consider a special class of surface operators
which preserve supersymmetry, in particular,
those supercharges which become BRST charges in the twisted theory.
Such surface operators can be defined on a more general embedded
surface $D$, which might be either closed or end on the boundary of $X$.
An example of this situation is a four-dimensional TQFT with corners,
shown on \linesfig,
which arisies when we consider a lift of a 3-manifold with boundary $\Sigma$
and line operators with end-points on $\Sigma$.

To summarize, including topological surface operators in
the four-dimensional gauge theory, we obtain a  functor from
the category of 3-manifolds with links and their cobordisms
to the category of graded vector spaces and homomorphisms:
\eqn\ykmapfunctor{ \CF (X;D) ~:~~ \CH_{Y;K} \to \CH_{Y';K'} }
Here, the knot homology $\CH_{Y;K}$ is the space of quantum ground states
in the four-dimensional gauge theory with surface operators and boundaries.
Similarly, the functor $\CF$ associates a number (the partition function)
to a closed 4-manifold with embedded surfaces,
and a category $\CF (\Sigma; p_i)$ to a surface $\Sigma$ with marked points, $p_i$,
which correspond to the end-points of the topological surface operators.

As in the theory without surface operators, the category $\CF (\Sigma; p_i)$
is either the category of A-branes or the category of B-branes on $\CM$,
depending on whether the topological reduction of the four-dimensional
gauge theory is A-model or B-model. Here, $\CM$ is the moduli space of
$\R^2$-invariant solutions in gauge theory on $X = \R^2 \times \Sigma$
with surface operators supported at $\R^2 \times p_i$.

\subsec{Braid Group Actions}

As we just explained, surface operators are the key ingredients for
realizing knot homologies in four-dimensional gauge theory.
Our next goal is to explain that every topological gauge theory
which admits surface operators is, in a sense, a factory that
produces examples of braid group actions on branes,
including some of the known examples as well as the new ones.\foot{It is worth
pointing out that, compared to \GWramified, where the braid group
action is associated with {\it local} singularities in the moduli space $\CM$,
in the present context the origin of the braid group action is associated
with {\it global} singularities.}

In general, the mapping class group of the surface $\Sigma$ acts on branes on $\CM$.
In particular, when $\Sigma$ is a plane with $n$ punctures,
the moduli space $\CM$ is fibered over the configuration
space ${\rm Conf}^n (\IC)$ of $n$ unordered points on $\IC$,
\eqn\conffib{\matrix{\CM \cr \downarrow \cr {\rm Conf}^n (\IC)}}
and the braid group $Br_n = \pi_1 ({\rm Conf}^n (\IC))$
(= the mapping class group of the $n$-punctured disk)
acts on the category $\CF (\Sigma)$.
Recall, that the braid group on $n$ strands, $Br_n$, has $n-1$
generators, $\sigma_i$, $i=1, \ldots, n-1$ which satisfy
the following relations
\eqn\braidgroup{\eqalign{
& \sigma_i \sigma_{i+1} \sigma_i = \sigma_{i+1} \sigma_i \sigma_{i+1} \cr
& \sigma_i \sigma_j = \sigma_j \sigma_i \quad, \quad\quad |i-j| >1
}}
where $\sigma_i$ can be reprsented by a braid with only one crossing
between the strands $i$ and $i+1$, as shown on the figure below.

\ifig\braidfig{A braid on four strands.}
{\epsfxsize2.0in\epsfbox{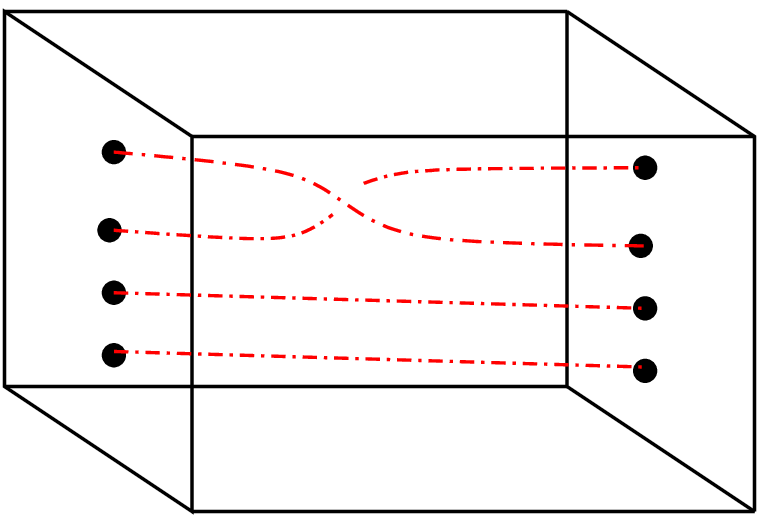}}

In gauge theory, the action of the braid group on branes is induced
by braiding of the surface operators. Namely, a braid, such as
the one on \braidfig, corresponds to a non-contactible loop in
the configuration space, ${\rm Conf}^n (\IC)$. As we go around
the loop, the fibration \conffib\ has a monodromy, which acts
on the category of branes $\CF (\Sigma)$ as an autoequivalence,
\eqn\braidauteq{\eqalign{ Br_n & \to {\rm Auteq} (\CF (\Sigma)) \cr
\beta & \mapsto \phi_{\beta} }}
The simplest situation where one finds the action of
the braid group $Br_n$ on A-branes (resp. B-branes)
on $\CM$ is when $\CM$ contans $A_{n-1}$ chain of Lagrangian
spheres (resp. spherical objects).

We remind that an $A_{n-1}$ chain of Lagrangian spheres is a collection
of Lagrangian spheres $\CL_1, \ldots, \CL_{n-1} \subset \CM$, such that
\eqn\lagrspheres{ \# \left( \CL_i \cap \CL_j \right)
=\cases{1 & $|i-j|=1$ \cr 0 & $|i-j| >1$ } }
These configurations occur when $\CM$ can be degenerated
into a manifold with singularity of type $A_{n-1}$.
Indeed, to any Lagrangian sphere $\CL \subset \CM$, one can
associate a symplectic automorphism of $\CM$,
the so-called generalized Dehn twist $T_{\CL}$ along $\CL$,
which acts on $H_* (\CM)$ as the Picard-Lefschetz monodromy
transformation
\eqn\PLefschetz{ (T_{\CL})_* (x) =
\cases{ x -  ([\CL] \cdot x) [\CL] & if $\dim (x) = \dim (\CL)$ \cr x & otherwise} }
As shown in \Seidel, Dehn twists $T_{\CL_i}$ along $A_{n-1}$ chains
of Lagrangian spheres satisfy the braid relations \braidgroup,
and this induces an action of the braid group with $n$ strands
on the category of A-branes, ${\rm\bf Fuk} (\CM)$.

The mirror of this construction gives an example of the braid
group action on B-branes \SThomas.
In this case, the braid group is generated by the twist functors
along spherical objects (``spherical B-branes'') which are mirror
to the Lagrangian spheres.
As the name suggests, an object $\CE \in D^b (\CM)$ is called
$d$-spherical if ${\rm Ext}^* (\CE , \CE)$ is isomorphic to
$H^* (S^d,\IC)$ for some $d>0$,
\eqn\sphericale{ {\rm Ext}^i (\CE , \CE) =
\cases{\IC & if $i=0$ or $d$ \cr 0 & otherwise} }
A spherical B-brane defines a twist functor
$T_{\CE} \in {\rm Auteq} (D^b (\CM))$
which, for any $\CF \in D^b (\CM)$, fits into exact triangle
\eqn\twisttriangle{
{\rm Hom}^* (\CE , \CF) \otimes \CE \longrightarrow \CF \longrightarrow T_{\CE} (\CF) }
where the first map is evaluation.
At the level of D-brane charges, the twist functor $T_{\CE}$ acts as,
{\it cf.} \PLefschetz,
$$
x \mapsto x + (v (\CE) \cdot x) ~v (\CE)
$$
where $v(\CE) = ch (\CE) \sqrt{Td (\CM)} \in H^* (\CM)$
is the D-brane charge (the Mukai vector) of $\CE$.

The mirror of an $A_{n-1}$ chain of Lagrangian spheres is
an $A_{n-1}$ chain of spherical objects, that is a collection
of spherical objects $\CE_1, \ldots, \CE_{n-1}$ which satisfy the
condition analogous to \lagrspheres,
\eqn\sphericalderived{
\sum_k \dim {\rm Ext}^k (\CE_i , \CE_j)
=\cases{1 & $|i-j|=1$ \cr 0 & $|i-j| >1$ } }
With some minor technical assumptions \SThomas,
the corresponding twist functors $T_{\CE_i}$ generate
an action of the braid group $Br_n$ on $D^b (\CM)$.
As we illustrate below, many examples of braid group actions
on branes can be found by studying gauge theory with surface operators.

\ifig\braidclosingfig{A particular brane $\tilde \CB$ which corresponds
to closing a braid on four strands.}
{\epsfxsize1.5in\epsfbox{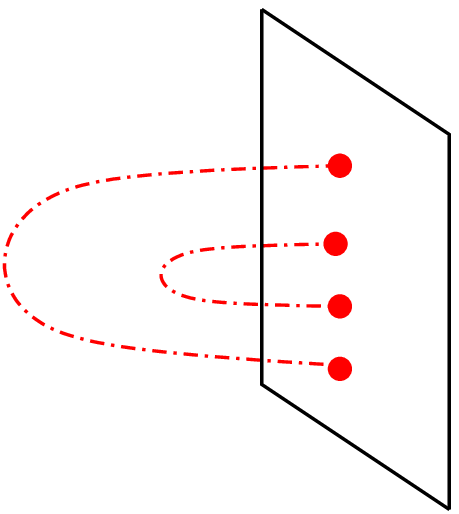}}

In A-model as well as in B-model, the braid group action on branes
can be used to write a convenient expression for knot homology, $\CH_K$,
of a knot $K$ represented as a braid closure.
Let $K$ be a knot obtained by closing a braid $\beta$ on both ends
as shown on \braidclosingfig.
Then, the space of quantum ground states, $\CH_K$, in the four-dimensional
gauge theory with a surface operator on $D = \R \times K$ can be represented
as the space of open string states between branes $\tilde \CB$
and $\tilde \CB' = \phi_{\beta} (\tilde \CB)$.
Here, $\tilde \CB$ is the basic brane which corresponds to the configuration
on \braidclosingfig, while $\tilde \CB'$ is the brane obtained from it by applying
the functor $\phi_{\beta}$; it corresponds to the braid $\beta$ closed on one side.
These are A-branes (resp. B-branes) in the case of A-model (resp. B-model),
and the space open strings is, {\it cf.} \abstrings,
\eqn\abbbstrings{ \CH_K 
= \cases{ HF^{{\rm symp}}_* (\CM; \tilde \CB, \phi_{\beta} (\tilde \CB)) & A-model \cr
{\rm Ext}^* (\tilde \CB , \phi_{\beta} (\tilde \CB)) & B-model} }
In particular, when topological reduction of the gauge theory gives A-model,
the branes $\tilde \CB$ and $\tilde \CB'$ are represented by Lagrangian submanifolds in $\CM$.
This leads to a construction of link homologies via symplectic geometry,
as in \refs{\SeidelSmith,\Manolescu,\Manolescusln}.


\newsec{Surface Operators and Knot Homologies in $\CN=2$ Gauge Theory}

Now, let us illustrate the general structures discussed in
the previous section in the context of $\CN=2$ topological
gauge theory in four dimensions.
For simplicity, we consider examples of $\CN=2$ gauge theories
with gauge groups $G=SU(2)$ and $G=U(1)$ known as
the Donaldson-Witten theory and the Seiberg-Witten theory, respectively.
In fact, these two theories are closely related \SWi\ --- the former describes
the low-energy physics of the latter --- and below we shall
use this fact to compare the corresponding knot homologies.


\subsec{Donaldson-Witten Theory and the Equivariant Knot Signature}

We start with pure $\CN=2$ super-Yang-Mills theory with gauge group $G$
which for simplicity we take to be $SU(2)$.
After the topological twist, the $\CN=2$ gauge theory can be formulated
on arbitrary 4-manifold $X$ and localizes on the anti-self-dual
(``instanton'') field configurations \WittenDonaldson:
\eqn\asdeqs{ F^+_A = 0 }
The space of quantum ground states on $\R \times Y$ is the instanton
Floer homology defined\foot{As in the original Floer's definition,
we mainly assume that $Y$ is a homology sphere when we talk about
$HF_*^{{\rm inst}} (Y)$ in order to avoid difficulties related to reducible connections.}
by studying the dradient flow of the Chern-Simons functional,
\eqn\hfinst{ \CH_Y = HF_*^{{\rm inst}} (Y) }
and the topological reduction \BJSV\
on $\R^2 \times \Sigma$ leads to a topological
A-model with the target space $\CM = \CM^G_{flat}$, the moduli space of
flat connections on $\Sigma$. As we already mentioned in the previous section
these facts, together with the interpretation of boundaries as D-branes,
naturally lead to the statement of the Atiyah-Floer conjecture \AFconj.
The Euler characteristic of $\CH_Y$ is the Casson invariant, $\la_G (Y)$,
which computes the Euler characteristic of the moduli space of flat
$G$-connections on $Y$,
\eqn\hycasson{ \chi( \CH_Y ) = 4 \la_G (Y) }
In the special case of $G=SU(2)$ that we are mainly considering here,
it is the standard Casson-Walker-Lescop invariant
$\lambda_{CWL} (Y)$ which sometimes we write simply as $\la (Y)$.

We note that, while the homological invariant $\CH_Y$
is difficult to study on 3-manifolds with $b_1 > 1$,
its Euler characteristic --- which is, at least formally,
computed by the partition function of the four-dimensional
gauge theory on $X = \S^1 \times Y$ --- is still given by
the Casson invariant \MMthreed,
\eqn\zdwy{ Z_{DW} (\S^1 \times Y) = 4 \la }
Since $b_2^+ (X) = b_1 (Y)$ for $X = \S^1 \times Y$,
computing \zdwy\ is much easier in the case $b_1 (Y) > 1$.
Indeed, in general the Donaldson-Witten partition function $Z_{DW} (X)$
can be written as a sum of the contribution of the Coulomb branch
(the $u$-plane integral) and two contributions, $Z_M$ and $Z_D$,
both of which are described by the Seiberg-Witten theory
(that we consider in more detail below):
\eqn\zzzz{ Z_{DW} = Z_u + Z_M + Z_D }
For manifolds $X = \S^1 \times Y$ with $b_1 (Y) > 1$ the $u$-plane
integral vanishes and we have $Z_M = Z_D = 2 \la (Y)$,
which then add up to \hycasson.
If $b_1 (Y) = 1$, the Donaldson-Witten partition function
$Z_{DW} (\S^1 \times Y)$ depends on the metric. In particular,
it should be compared with the Euler characteristic of $\CH_Y$
in the chamber $R \to \infty$, where $R$ is the radius of $\S^1$.
In this case, the $u$-plane integral is non-zero,
and instead of \zdwy\ one finds a similar expression with
the ``correction'' $- {4 \over 6} | {\rm Tor} H_1 (Y,\Z) |$,
see \MMthreed\ for more details.

\medskip\noindent{{\it Surface Operators}}\medskip

Now let us consider surface operators in the Donaldson-Witten
theory which correspond to the singularity of the gauge field
of the form
\eqn\asingularity{ A = \a d \theta + \ldots }
Here, $(r,\theta)$ are radial coordinates in the normal plane,
$\a$ is the parameter which labels surface operators and takes values in
$\frak t = {\rm Lie} (\TT)$, the Lie algebra of the maximal torus $\TT \subset G$,
and the dots in \asingularity\ stand for less singular terms.
More precisely, inequivalent choices of $\a$  are labeled by
elements in $\TT = \frak t / \Lambda_{{\rm cochar}}$ since gauge
transformations shift $\a$ by vectors in the cocharacter lattice
$\Lambda_{{\rm cochar}}$ of $G$.
For example, for $G=SU(2)$ we have $\TT = U(1)$.

In the presence of a surface operator on $D \subset X$,
supersymmetric field configurations in this theory
are described by the instanton equations \asdeqs:
\eqn\asdsurface{ F^+_A = 2\pi \a (\delta_D)_+ }
perturbed by the term $2\pi \a (\delta_D)_+$,
where $(\delta_{D})_+$ denotes the self-dual part of the
cohomology class that is Poincar\'e dual to the surface $D$.
In the context of $SU(2)$ gauge theory, such surface operators
were extensively used in the work of Kronheimer and Mrowka on minimal
genus problems of embedded surfaces in 4-manifolds \refs{\KMi,\KMii}.

According to the general rules outlined in the previous section,
to a 4-manifold $X = \R \times Y$ and a surface operator on $D = \R \times K$
labeled by $e^{2 \pi i \a} \in \TT$ the Donaldson-Witten theory
associates a vector space, the space of quantum ground states,
\eqn\hykinst{ \CH_{Y;K,\a} = HF_{*}^{{\rm inst}} (Y;K,\a) }
Just like the ordinary instanton Floer homology \hycasson,
it categorifies a Casson-like invariant,
\eqn\hykcasson{ \chi (\CH_{Y;K,\a}) = \la_{\a} (Y;K) }
which counts flat $SU(2)$ connections on a homology sphere $Y$
with the prescribed singularity \asingularity\ along $K$.

In order to describe $\la_{\a} (Y;K)$ more explicitly,
it is convenient to decompose $Y$ as in \heegard\
into a tubular neighborhood of the knot $K$, $Y_1 = N(K)$,
and its complement, $Y_2 = Y \setminus N(K)$,
glued along the common boundary $\Sigma \cong T^2$.
As we already mentioned earlier, topological reduction
of the Donaldson-Witten theory on $\Sigma$ yields
a topological A-model with the target space $\CM = \CM_{flat}^G$,
the moduli space of flat connections on $\Sigma$:
\eqn\mgflatrep{ \CM_{flat}^G = \{ \rho :~ \pi_1 (\Sigma) \to G \} / {\rm conj.} }
For $\Sigma = T^2$ this moduli space is the quotient,
$\CM_{flat}^G = (\TT \times \TT) / \Weyl$,
of two copies of the maximal torus by the Weyl group of $G$.
In particular, for $G=SU(2)$ the corresponding moduli space
$\CM_{flat}^G \cong T^2 / \Z_2$ is often called the ``pillowcase''.
Similarly, each component in the decomposition $Y = Y_1 \cup_{\Sigma} Y_2$
defines an A-brane supported on a Lagrangian submanifold in $\CM_{flat}^G$.
If we denote Lagrangian submanifolds associated to $Y_1$ and $Y_2$,
respectively, by $\CL_{\a}$ and $\CL_{Y \setminus K}$,
then the invariant \hykcasson\ is given by their
intersection number (in the smooth part of $\CM_{flat}^G$):
\eqn\cassonlinint{
\la_{\a} (Y;K) = \# \left( \CL_{\a} \cap \CL_{Y \setminus K} \right) }
Note, the Lagrangian brane supported on $\CL_{\a}$ does not
depend on $K$ or $Y$, while the Lagrangian brane supported
on $\CL_{Y \setminus K}$ does not depend on $\a$.
Indeed, $\CL_{\a}$ is simply the set
of representations $\rho \in \CM_{flat}^G$ taking the meridian
of the knot $K$ to a matrix of trace $\tr( \rho_{\mu} ) = 2 \cos \pi \a$.
Similarly, the Lagrangian brane supported on $\CL_{Y \setminus K}$
corresponds to flat connections on $\Sigma = T^2$ which
can be extended to flat connections on $Y \setminus K$.
In other words, $\CL_{Y \setminus K}$ is the image of
$\CM_{flat}^G (Y \setminus K)$ under the restriction map
$$
r: \CM_{flat}^G (Y \setminus K) \to \CM_{flat}^G (T^2)
$$
induced by the inclusion of the torus boundary of the knot
complement, $T^2 \hookrightarrow Y \setminus K$.

To summarize, surface operators in the Donaldson-Witten theory
lead to a variant of the instanton Floer homology, $\CH_{Y;K.\a}$,
whose Euler characteristic is a Casson-like invariant \cassonlinint.
This is precisely the definition of the knot invariant which was
introduced and studied in \refs{\Lin,\CLMillerCasson,\Herald}
(see also \refs{\Heraldi,\Kroll}).
This invariant, sometimes called Casson-Lin invariant, is well-defined
away from the roots of the Alexander polynomial of $K$ and turns out to
be equal to the linear combination of more familiar invariants, $\a \in [0,1]$,
\eqn\cassonlinsign{
\lambda_{\a} (Y;K) = 4 \lambda (Y) + \half \sigma_{\a} (K) }
where $\lambda (Y)$ is the Casson invariant of $Y$ and
$\sigma_{\a} (K) : U(1) \to \Z$ is the equivariant signature function
({\it a.k.a.} Levine-Tristram signature) of the knot $K$.
Homology theory categorifying $\lambda_{\a} (Y;K)$
was constructed in \CollinSteer\ (see also \refs{\Collin,\Collinii})
and, therefore, is expected to be the same as \hykinst.

We remind that, for a knot $K$ in a homology sphere $Y$,
the normalized Alexander polynomial is defined as
\eqn\alexdef{
\Delta (K;q) = \det \left( q^{1/2} {\bf V} - q^{-1/2} {\bf V}^T \right) }
where ${\bf V}$ is the Seifert matrix of $K$ and $q= e^{2 \pi i\a}$.
Note, that $\Delta (K;q) = \Delta (K;q^{-1})$.
The equivariant signature $\sigma_{\a} (K)$
is defined as the signature of the Hermitian matrix
\eqn\bkdef{
B_K (q) = (1-q) {\bf V} + (1 - \bar q) {\bf V}^{T} }
The equivariant signature function changes its value only
if $q= e^{2 \pi i\a}$ is a root of the Alexander polynomial.
It vanishes for $\a$ near $0$ or $1$,
\eqn\eqsignlim{ \lim_{\a \to 0,1} \sigma_{\a} (K) = 0 }
and equals the standard knot signature, $\sigma (K)$, for $\a = {1 \over 2}$.
In particular, for $Y = \S^3$ and $\a= {1 \over 2}$
we get the original Lin's invariant \Lin\ and the corresponding
homology theory categorifying $\lambda_{{1 \over 2}} (Y;K)$
was constructed --- as symplectic Floer homology \abbbstrings\
of the braid representative of $K$ --- in \LicassonL.

\subsec{Seiberg-Witten Theory}

Now let us consider $\CN=2$ twisted gauge theory
with abelian gauge group $G=U(1)$ coupled to a single monopole field $M$.
This theory localizes on the solutions to
the Seiberg-Witten equations for abelian monopoles \Wittenmonopoles:
\eqn\sweqs{\eqalign{
& F^+_A + i (\bar M M)_+ = 0 \cr
& \D_A M = 0 }}
which follow from the $\CN=2$ topological gauge theory\foot{
Up to the finite group $H^1 (X,\Z_2)$,
the set of Spin$^c$ structures on a 4-manifold $X$ is
parameterized by integral cohomology classes
which reduce to $w_2 (X)$ mod 2,
$$
{\rm Spin}^c (X) = \{ x \in H^2 (X,\Z) ~\vert~ x \equiv w_2 (X) ~{\rm mod}~2 \}
$$
Given a Spin$^c$ structure $x \in {\rm Spin}^c (X)$, let $L$ be
the corresponding Hermitian line bundle, and $S^{\pm}_L$ the
corresponding spinor bundles. Then, the Seiberg-Witten monopole
equations \sweqs\ are equations for a pair $(A,M)$,
where $A$ is a unitary connection on $L$ and $M$
is a smooth section of $S^+_L$.
In writing the equations \sweqs\ we used the Dirac operator,
$\D_A : S^+_L \to S^-_L$, and a map
$$
\eqalign{
\Omega^0 (S^+_L) & \to \Omega^0 ( {\rm ad}_0 S^+_L ) \cr
M & \mapsto i (\bar M M)_+ }
$$
where ${\rm ad}_0 S^+_L \cong \Lambda ^2_+$ is the subbundle
of the adjoint bundle of $S^+_L$ consisting of the traceless
skew-Hermitian endomorphisms, which can be identification with
the space of self-dual 2-forms.}.
The partition function of this theory on a 4-manifold $X$
with 2-observables included is a generating function of
the Seiberg-Witten invariants, $SW_X (x)$, which ``count''
solutions to \sweqs\ and can be viewed as a function
of $x \in {\rm Spin}^c (X)$,
$$
SW_X : {\rm Spin}^c (X) \to \Z
$$
To be more precise, the Seiberg-Witten invariants are defined as
integrals over $\CM_x$, the moduli space of solutions to
the Seiberg-Witten equations \sweqs,
$$
SW_X (x) = \int_{\CM_x} a_D^{d_x/2}
$$
where $d_x = {1 \over 4} (x^2 - 2 \chi (X) - 3 \sigma (X))$
is the virtual dimension of $\CM_x$,
and $a_D$ is a 2-form which represents the first Chern class
of the universal line bundle on the moduli space $\CM_x$.

The space of quantum ground states in this theory is
the Seiberg-Witten monopole homology,
\eqn\hyhmy{ \CH_{Y} = HM_* (Y) }
which is conjectured to be isomorphic to the Heegard Floer homology,
see {\it e.g.} \KMbook:
\eqn\monhfconj{ HM_* (Y) \cong HF_* (Y) }
In turn, the Heegard Floer homology $HF_* (Y)$  --- as well as its analog for
knots, the knot Floer homology $HFK_* (K)$, that is closer to our interest ---
is defined as the symplectic Floer homology of certain Lagrangian
submanifolds in the symmetric product space of the form \refs{\OShfk,\Rasmussen,\OShf},
\eqn\symmprodk{ \CM = {\rm Sym}^k (\Sigma) }
The symmetric product space and Lagrangian submanifolds in it
naturally appear in the topological reduction of the Seiberg-Witten theory.
Indeed, on $X = \R^2 \times \Sigma$ the equations \sweqs\
reduce to the vortex equations in the abelian Higgs model,
and the moduli space of solutions to these equations,
namely the moduli space of charge $k$ vortices,
is the symmetric product space \symmprodk, see \JaffeT.
As in the case of the Donaldson-Witten theory,
the topological reduction of the Seiberg-Witten theory
leads to the topological A-model\foot{In fact, this is true
for any four-dimensional $\CN=2$ gauge theory with the same
type of topological twist \BJSV.} with $\CM$ as the target space,
and the corresponding category of branes in this case
is the category of A-branes,
\eqn\fukayam{ \CF (\Sigma) = {\rm\bf Fuk} (\CM) }

According to the general rules explained in the previous section,
the Euler characteristic of the homology theory \hyhmy\ - \monhfconj\
is given by the partition function of the Seiberg-Witten theory
on $X = \S^1 \times Y$ (in the chamber $R \to \infty$):
\eqn\zswynoq{ Z_{SW} (Y) = \sum_{x \in {\rm Spin}^c (Y)} SW_Y (x) }
If $b_1 (Y) > 1$, then there are no wall-crossing phenomena and $Z_{SW} (Y)$
can be equivalently viewed as the partition function of the three-dimensional
gauge theory on $Y$ obtained by the dimensional reduction of the Seiberg-Witten theory.
For a fairly general class of 3-manifolds $Y$, the partition function
\zswynoq\ is equal to the Casson invariant of $Y$, {\it c.f.} \zdwy:
\eqn\zswcasson{ Z_{SW} (Y) = \lambda (Y) }
For instance, for 3-manifolds with $b_1 (Y) > 1$ it follows {\it e.g.} from
the general result of Meng and Taubes \MengT\ that will be discussed
in more detail below.
On the other hand, for homology spheres the definition of
the Seiberg-Witten invariants requires extra care.
However, once this is done, one can show that \zswcasson\
still holds for suitably defined $Z_{SW} (Y)$;
see \refs{\Lim,\CLMiller,\Nicolaescu} for a mathematical proof
and \BTonRW\ for a physical argument based on the duality
with Rozansky-Witten theory \RW.

\medskip\noindent{{\it Surface Operators}}\medskip

As in the Donaldson-Witten theory, we can introduce surface operators
by requiring the gauge field to have the singularity of the form \asingularity.
In the Seiberg-Witten theory, such surface operators are labeled
by  $e^{2 \pi i \a} \in U(1)$.
In the presence of a surface operator on $D \subset X$,
supersymmetric field configurations are described by
the perturbed Seiberg-Witten monopole equations, {\it cf.} \refs{\Wittenmonopoles,\Taubesbook}:
\eqn\sweqs{\eqalign{
& F^+_A + i (\bar M M)_+ = 2\pi \a (\delta_D)_+ \cr
& \D_A M = 0 }}
As usual, in order to obtain a homological invariant of a knot $K$
in a 3-manifold $Y_0$ one should consider the Hilbert space of
the gauge theory on $X = \R \times Y_0$ with a surface operator
on $D = \R \times K$.
In the context of Seiberg-Witten theory, this gives a vector space
$\CH_{Y_0;K;x,\a}$.
More generally, given a link $L$ with $\ell$ components one can
introduce $\ell$ surface operators, each with its own parameter $\a_i$,
$i = 1, \ldots, \ell$. The corresponding Hilbert space is
\eqn\hhm{ \CH_{Y_0; L; x_i, \a_i} = HM_* (Y; x_i, \a_i) }
where $Y = Y_0 \setminus L$ is the link complement,
and parameters $\a_i$ determine the boundary conditions on $Y$.
Namely, the holonomy of the $U(1)$ gauge connection $A$ along
the meridian of the $i$-th link component should be equal to $e^{2\pi i \a_i}$.
We will be mainly interested in the case where $Y_0 = \S^3$
and $Y = \S^3 \setminus L$ is the link complement.
In this case, \hhm\ gives $(\ell + 1)$-graded link homology.

We introduce the graded Euler characteristic of the homological invariant \hhm,
\eqn\hhmeuler{ \tau_{\a_i} (Y;q_i) :=
\sum_{x \in H(Y)} \chi \left( HM_* (Y; x_i, \a_i) \right) \cdot q^x
}
which is a formal power series in $q_i^{\pm 1}$, where $q_i = e^{h_i}$
and $h_i$ are the generators of a free abelian group,
$$
H (Y) := H_1 (Y,\Z) / {\rm torsion} \cong \Z^{b_1 (Y)}
$$
In particular, when $Y$ is a link complement,
the group $H(Y) = H_1 (Y,\Z) \cong \Z^{\ell}$
is generated by the meridians of the link components.

In general, $\tau_{\a_i} (Y;q_i)$ is a non-trivial function
of $q_i$ and $\a_i$.
It is equal to the partition function of the Seiberg-Witten
theory on $X = \S^1 \times Y$ (in the chamber $R \to \infty$):
\eqn\sweuler{
\tau_{\a_i} (Y;q_i) = \sum_{x \in H(Y)} SW_Y (x,\a_i) \cdot q^x }
This function is an interesting generalization of
the Reidemeister-Milnor torsion, on the one hand,
and the equivariant knot signature, on the other.
Indeed, since the Seiberg-Witten theory is the low-energy description
of the Donaldson-Witten theory,
we expect the relation to the equivariant knot signature.
On the other hand,
if $\a_i$ is near $0$ or $1$ for all $i = 1, \ldots, \ell$, as in \eqsignlim,
then the partition function $Z_{SW} (Y; x_i, \a_i) = \tau_{\a_i} (Y;q_i)$
becomes the ordinary partition function of the link complement $Y$
studied by Meng and Taubes \MengT\ who showed that it is equal
to the Reidemeister-Milnor torsion. Hence,
\eqn\zswlim{ \lim_{\a_i \to 0,1} \tau_{\a_i} (Y;q_i) = \tau (Y;q_i) }
where $\tau (Y;q_i)$ is the ordinary Reidemeister-Milnor torsion of $Y$.
In particular, for $b_1 (Y) > 1$ we have $\tau (Y;q_i) = \Delta (L;q_i)$,
so that in this limit the homological invariant \hhm\ categorifies
the multi-variable Alexander polynomial $\Delta (L;q_i)$ of the link $L$,
\eqn\hmalex{
\sum_{x \in H(Y)} \chi \left( \CH_{L; x} \right) \cdot q^x = \Delta (L;q) }
This suggests to identify the $(\ell + 1)$-graded homology theory \hhm\
with the link Floer homology \OShfl,
\eqn\hmhfl{ \CH_{L; x} = HFL_* (L;x) }
In the case of knots, the relation to the Alexander polynomial $\Delta (K;q)$
is slightly more delicate, in part due to metric dependence and wall crossing.
It turns out, however, that even though individual Seiberg-Witten invariants
are different in the positive and negative chamber, the corresponding
generating functions are both equal to the Milnor torsion \MengT,
so that \zswlim\ still holds.
Note, that specializing \zswlim\ further to $q_i = 1$, we recover \zswcasson.
It would be interesting to study the invariant $\tau_{\a} (Y;q)$ further,
in particular, its relation to the equivariant knot signature $\sigma_{\a} (K)$.


\newsec{Surface Operators and Knot Homologies in $\CN=4$ Gauge Theory}

Now, let us consider surface operators and knot homologies
in the context of $\CN=4$ topological super-Yang-Mills theory in four dimensions.
Specifically, we shall consider the GL twist of the theory \KWitten,
with surface operators labeled by regular semi-simple conjugacy classes \GWramified.
As we shall explain below, this theory provides a natural framework
for categorification of the $\CCG$ Casson invariant, which counts
flat connections of the complexified gauge group $\CCG$.

The topological reduction of this theory leads to a $\CN=4$ sigma-model
\refs{\BJSV,\HMStrominger,\KWitten},
whose target space is a hyper-Kahler manifold $\CM_H (\Sigma, G)$,
the moduli space of solutions to the Hitchin equations on $\Sigma$ \Hitchin:
\eqn\hiteqs{\eqalign{
& F - \phi \wedge \phi = 0 \cr
& d_A \phi = 0 \quad,\quad d_A \star \phi = 0
}}

This twist of the $\CN=4$ super-Yang-Mills theory has a rich spectrum
of supersymmetric surface operators. In particular, here we will be
interested in the most basic type of surface operators, which correspond
to the singular behavior of the gauge field $A$ and the Higgs field $\phi$
of the form \GWramified:
\eqn\tamesurf{\eqalign{
& A = \a d \theta + \ldots, \cr
& \phi = \beta {dr \over r} - \gamma d \theta + \ldots
}}
where $\a,\b,\g \in {\frak t}$, and the dots stand for the terms less singular at $r=0$.
For generic values of the parameters $\a,\b,\g$, eq. \tamesurf\ defines a surface
operator associated with the regular semi-simple conjugacy class $\frak C \in \CCG$.

According to the general rules explained in section 2,
this topological field theory associates a homological invariant $\CH_Y$
to a closed 3-manifold $Y$ and, more generally, a knot homology $\CH_{Y;K}$
to a 3-manifold with a knot (link) $K$.
These homologies can be computed as in \abstrings\ and \abbbstrings\
using the Heegard decomposition of $Y$ as well as the braid group action on branes.
The branes in questions\foot{{\it e.g.} branes $\CB_1$ and $\CB_2$
associated with the Heegard decomposition $Y = Y_1 \cup_{\Sigma} Y_2$}
are branes of type $(A,B,A)$ with respect to the three complex structures $(I,J,K)$
of the hyper-Kahler space $\CM_H (\Sigma, G)$.
We can use this fact and analyze the branes in different complex structures
in order to gain a better understanding of the homological invariant $\CH_{Y,K}$
as well as the $\CCG$ Casson invariant itself.
For example, in complex structure $I$ it corresponds to counting
parabolic Higgs bundles, a fact that has already been used {\it e.g.} in \BCurtis\
for studying the $SL(2,\IC)$ Casson invariant for Seifert fibered homology spheres.

\medskip\noindent{{\it Complex Structure $J$: Counting Flat Connections}}\medskip

The B-model in complex structure $J$ is obtained, {\it e.g.} by setting
the theta angle to zero, $\r (\tau) = 0$, and choosing $t=i$
(where $t$ is a complex parameter that labels
a family of GL twists of the $\CN=4$ super-Yang-Mills \KWitten).
In complex structure $J$, the moduli space
$\CM_H (\Sigma, G) \cong \CM_{flat}^{\CCG} (\Sigma)$
is the space of complexified flat connections $\CA = A + i \phi$,
and the surface operator \tamesurf\ creates a holonomy,
$$
V = {\rm Hol} (\CA),
$$
which is conjugate to $\exp (- 2\pi (\a - i\g))$.
Furthermore, at $t=i$ the supersymmetry equations of
the four-dimensional gauge theory are equivalent
to the flatness equations, $d \CA + \CA \wedge \CA = 0$,
which explains why (from the viewpoint of complex structure $J$)
the partition function of this theory on $X = \S^1 \times Y$
with a surface operator on $D = \S^1 \times K$
computes the $\CCG$ Casson invariant,
$$
Z = \la_{\CCG} (Y;K)
$$
The space of ground states, $\CH_{Y;K}$,
is a categorification of $\la_{\CCG} (Y;K)$.
In general, both $\la_{\CCG} (Y;K)$ and $\CH_{Y;K}$ depend on the holonomy $V$,
which characterizes surface operators.
However, if $V$ is regular semi-simple, as we consider here,
then  $\la_{\CCG} (Y;K)$ and $\CH_{Y;K}$ do not depend on
a particular choice of $V$.

\medskip\noindent{{\it Complex Structure $K$}}\medskip

Since the four-dimensional topological gauge theory
(even with surface operators)
does not depend on the parameter $t$ that labels different twists,
we can take $t=1$, which leads to the A-model on $\CM_H (\Sigma, G)$
with symplectic structure $\omega_K$.
This theory computes the same $\CCG$ Casson invariant
and its categorification, $\CH_{Y;K}$, but via counting
solutions to the following equations on $Y$ \KWitten:
\eqn\akthreed{\eqalign{
& F - \phi \wedge \phi = \star \big( D \phi_0 - [A_0 , \phi]\big) \cr
& \star D \phi = [\phi_0, \phi] + D A_0 \cr
& \star D \star \phi + [A_0 , \phi_0] = 0
}}
rather than flat $\CCG$ connections.
In particular, given a Heegard decomposition $Y = Y_1 \cup_{\Sigma} Y_2$,
the space of solutions to the equations \akthreed\ on $Y_1$ (resp. $Y_2$)
defines a Lagrangian A-brane in $\CM_H (\Sigma, G)$ with respect to $\omega_K$.
This allows to express $\CH_{Y;K}$ as the space of open string states
between the corresponding A-branes $\CB_1$ and $\CB_2$, {\it cf.} \abstrings,
$$
\CH_{Y;K} = HF^{{\rm symp}}_* (\CM_H; \CB_{1}, \CB_{2})
$$
This alternative definition of the $\CCG$ Casson invariant and
its categorification that follows from the twisted $\CN=4$
gauge theory can be useful, for instance, for understanding
situations when the $(A,B,A)$ branes $\CB_1$ and $\CB_2$
intersect at singular points in $\CM_H$ or over higher-dimensional subvarieties.

\medskip\noindent{{\it Categorification of the $SL(2,\IC)$ Casson Invariant}\medskip

Now, let us return to the complex structure $J$ and, for simplicity,
take the gauge group to be $G=SU(2)$.
Furthermore, we shall consider an important example
of the sphere with four punctures:
$$
\Sigma = \cp^1 \setminus \{ p_1, p_2, p_3, p_4 \}
$$
which in gauge theory corresponds to inserting four surface operators.
In complex structure $J$, $\CM_H (\Sigma, G)$ is the moduli space
of flat $\CCG = SL(2,\IC)$ connections with fixed conjugacy class
of the monodromy around each puncture.
It can be identified with the space of conjugacy classes of monodromy representations
$$
\CM_H (\Sigma, G) \cong
\{ \rho :~ \pi_1 (\Sigma) \to \CCG ~\vert~ \rho (\gamma_i) \in \frak C_i \} / \sim
$$
where the representations are restricted to take the simple loop $\gamma_i$
around the $i$-th puncture into the conjugacy class $\frak C_i \subset \CCG$.

\ifig\fourptfig{Sphere with four punctures.}
{\epsfxsize1.2in\epsfbox{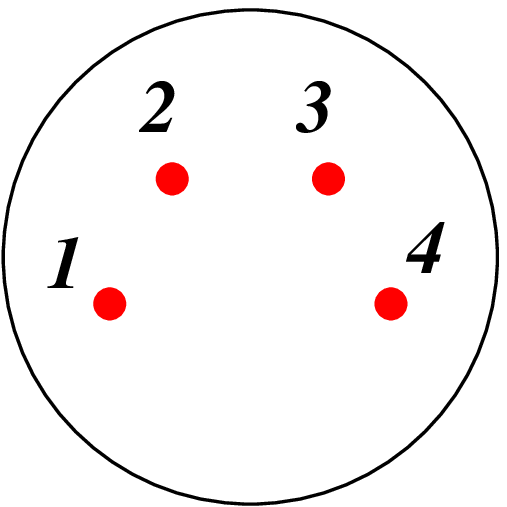}}

Using the fact that $\pi_1 (\Sigma)$ is free on three generators,
we can explicitly describe the moduli space $\CM_H (\Sigma, G)$ by introducing
holonomies of the flat $SL(2,\IC)$ connection around each puncture,
\eqn\misdef{
V_i = {\rm Hol}_{p_i} (\CA) \quad, \quad i=1, \ldots, 4 }
where $V_1 V_2 V_3 V_4 = 1$ and each $V_i$ is in a fixed conjugacy class.
Following \refs{\Iwasaki,\IISaito,\IwasakiU,\Sakai,\HitSchles},
we introduce the local monodromy data
\eqn\aisdef{
a_i = \cases{ \tr V_i & $i=1,2,3$ \cr \tr (V_3 V_2 V_1) & $i=4$} }
and
\eqn\thetasdef{\eqalign{
& \theta_1 = a_1 a_4 + a_2 a_3 \cr & \theta_2 = a_2 a_4
+ a_1 a_3 \cr & \theta_3 = a_3 a_4 + a_2 a_1 \cr & \theta_4 = a_1
a_2 a_3 a_4 + \sum_{i=1}^4 a_i^2 - 4 }}
which determines the conjugacy classes of $V_i$.
We also introduce the variables
\eqn\xmm{\eqalign{
& x_1 = \tr (V_3 V_2) \cr
& x_2 = \tr (V_1 V_3) \cr
& x_3 = \tr (V_2 V_1)
}}
which will be the coordinates on the moduli space $\CM_H (\Sigma, G)$.
Namely, the moduli space we are interested in is
$$
\CM_H (\Sigma, G) = \{ (V_1,\ldots,V_4) ~\vert~ V_i \in \frak C_i, ~V_1 V_2 V_3 V_4 = 1 \} / \CCG
$$
In terms of the variables \xmm, it can be explicitly described
as the affine cubic
\eqn\mcubic{
\CM_H (\Sigma, G) = \{ (x_1,x_2,x_3) \in \IC^3  ~\vert~ f(x_i,\theta_m) = 0 \} }
where
\eqn\fcubic{
f(x_i,\theta_m) = x_1 x_2 x_3 + \sum_{i=1}^3 (x_i^2 - \theta_i x_i) + \theta_4 }
%

\medskip\noindent{{\it Singularities in $\CM_H$}}\medskip

For certain values of the monodromy data, the moduli space $\CM_H$
becomes singular. It is important to understand the nature of
the singularities and when they develop. In fact, as we shall see
below, inetersting examples of branes pass through such singularities.

The descriminant $\Delta(f)$ of the cubic \fcubic\
is a polynomial in $a_i$ of total degree $16$ \Iwasakid:
\eqn\dcubic{ \Delta (a)
= \left( \prod_{\epsilon_1 \epsilon_2 \epsilon_3 = 1}
\big(a_4 + \sum \epsilon_i a_i \big)
- \prod_{i=1}^3 (a_i a_4 - a_j a_k)
\right)^2 \prod_{i=1}^4 (a_i^2 - 4) }
where $\epsilon_i = \pm 1$.
A special subfamily of cubics \fcubic, which will play an important role
in applications to knot invariants discussed below, corresponds to the case
where all monodromy parameters $a_i$ are equal, $a_i=a$, $i=1,2,3,4$.
In this case,
\eqn\thetaequala{\eqalign{
& \theta_i = 2a^2 \quad\quad,\quad\quad i=1,2,3 \cr
& \theta_4 = a^4 + 4a^2 - 4 }}
and it is easy to verify that $\Delta (a)=0$. Specifically,
for generic values of the parameter $a$, the moduli space $\CM_H$
has three simple singularities of type $A_1$ (double points) at
\eqn\xxxaone{
(x_i,x_j,x_k) = (a^2-2,2,2) }
These singularities correspond to reducible flat connections.
For special values of $a$, the singularities can become worse
and/or additional singularities can appear. For example,
for $a^2=0$ a new singularity of type $A_1$ develops at
the point $(x_1,x_2,x_3) = (-2,-2,-2)$. On the other hand,
for $a^2=4$ the moduli space has a simple singularity of
type $D_4$ at $(x_1,x_2,x_3) = (2,2,2)$.

\medskip\noindent{{\it Braid Group Action}}\medskip

The mapping class group of $\Sigma$, which in the present case is
the braid group $Br_3$, acts on the family of cubic surfaces \mcubic\
by polynomial automorphisms. In particular, one can verify that
the generators $\sigma_i$, $i=1,2,3$, represented as \Iwasaki:
\eqn\broncubic{
\sigma_i ~:~~ (x_i,x_j,x_k,\theta_i,\theta_j,\theta_k,\theta_4)
\to (\theta_j - x_j - x_k x_i, x_i, x_k, \theta_j,\theta_i,\theta_k,\theta_4) }
satisfy the relations $\sigma_i \sigma_j \sigma_i = \sigma_j \sigma_i \sigma_j$
and $\sigma_k = \sigma_i \sigma_j \sigma_i^{-1}$. Here and below
we denote by $(i,j,k)$ any cyclic permutation of $(1,2,3)$.

\medskip\noindent{{\it Examples}}\medskip

Let us consider examples of $(A,B,A)$ branes that arise from knotted
surface operators in $\R \times \B^3$, where $\B^3$ denotes a 3-dimensional ball.
We consider surface operators which are extended along the $\R$ direction
and which meet the boundary $\S^2 = \p \B^3$ at four points.
The simplest example of such brane is
\eqn\albrane{
{\rm brane}~\tilde \CB~=~~~
{\lower0.55in \hbox{\epsfxsize0.9in\epsfbox{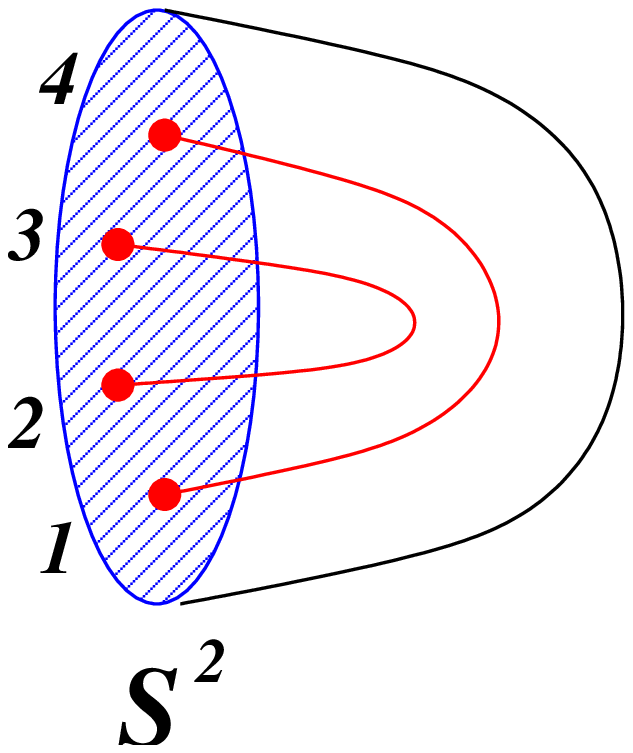}}} }
We shall denote this brane $\tilde \CB$ (or $\tilde \CB_{(14)(23)}$
if we wish to specify which pairs of points on $\S^2$ it connects).
Since the brane \albrane\ identifies the monodromies around
the points 1 and 4 (resp. 2 and 3),
\eqn\mmmm{V_1 = V_4^{-1} \quad\quad,\quad\quad V_2 = V_3^{-1}}
it can be explicitly described as a subvariety of $\CM_H$ defined by
\eqn\xonebrane{ x_1 = \tr (V_3 V_2) = 2 }
Of course, we also need to set $a_1 = a_4$ and $a_2 = a_3$,
so that
\eqn\thonebrane{\eqalign{
& \theta_1 = a_1^2 + a_2^2 \cr
& \theta_2 = \theta_3 = 2 a_1 a_2 \cr
& \theta_4 = a_1^2 a_2^2 + 2a_1^2 + 2a_2^2 - 4 }}
Substituting \xonebrane\ and \thonebrane\ into the cubic equation
$f(x_i,\theta_m)=0$, we find that the brane \albrane\
can be described as a degenerate quadric,
\eqn\albranequadric{ (x_2 + x_3 - a_1 a_2)^2 =0 }
One can also think of it as a set of two coincident branes on $x_2 + x_3 = a_1 a_2$.
By acting on this brane with the elements of the braid
group \broncubic, we can construct other examples of $(A,B,A)$ branes in $\CM_H$.
Furthermore, by closing the braid one can obtain homological
invariants of knots (links) in $\S^3$ as spaces of open strings between two such branes.
In the rest of this section, we consider a few explicit examples.

\medskip\noindent
$\underline{{\rm Unknot:}}$ One way to construct the unknot is to take
surface operators which correspond to two branes of type \albrane,
as shown on the figure below:

\ifig\unknotaa{Unknot in $\S^3$ can be represented as a union of two branes $\tilde \CB$.}
{\epsfxsize1.5in\epsfbox{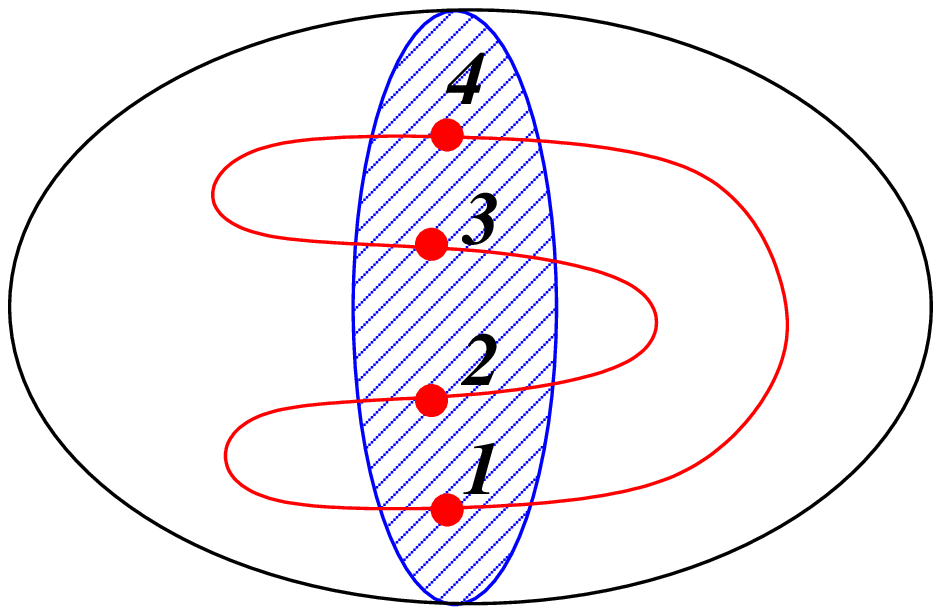}}

\noindent
The two branes on this figure are branes $\tilde \CB_{(14)(23)}$
and $\tilde \CB_{(12)(34)}$. We already discussed the first brane:
it is described by the conditions \xonebrane\ -- \albranequadric.
Similarly, the brane $\tilde \CB_{(12)(34)}$ is given by $V_1 = V_2^{-1}$,
which implies $V_3 = V_4^{-1}$,
\eqn\xthreebrane{ x_3 = \tr (V_2 V_1) = 2 }
and the corresponding conditions for $\theta_i$, {\it cf.} \thonebrane.
Altogether, the conditions describing these two branes imply
that the local monodromy data should be identified,
\eqn\aaaaa{a_1 = a_2 = a_3 = a_4 = a}
This condition is very natural, of course, and will be relevant in
all the examples where the resulting link has only one connected component,
{\it i.e.} is actually a knot.
Furthermore, for the unknot in \unknotaa\ we have:
\eqn\mmmmunknot{ V_1 = V_2^{-1} = V_3 = V_4^{-1} }
These equations describe the intersection points of branes $\tilde \CB_{(14)(23)}$
and $\tilde \CB_{(12)(34)}$. Using \xonebrane\ and \xthreebrane, it is easy to see
that there is only one such point (of multiplicity 2):
\eqn\xxxunknotaa{ (x_1,x_2,x_3) = (2,a^2-2,2) }
This is precisely one of the singular points \xxxaone\
where the moduli space $\CM_H$ has $A_1$ singularity
(for generic values of $a$) due to reducible representations.
Therefore, we conclude that the cohomology of the unknot, $\CH^{sl(2)}_{{\rm unknot}}$,
is given by the space of open string states for two different branes
intersecting at the $A_1$ singularity in $\CM_H$.
We point out that the values of $x_i$ in \xxxunknotaa\
can be read off directly from \unknotaa.
Indeed, $x_1 = 2$ simply follows from the fact that the combined monodromy
around the points 2 and 3 is equal to the identity (similarly for $x_3=2$).
In order to explain $x_2 = a^2-2$, it is convenient to introduce
the eigenvalues $m^{\pm 1}$ of the monodromy matrix $V_1$.
Of course, $m$ is related to the local monodromy parameter $a$,
namely $a = m + m^{-1}$. Moreover, since $V_1 = V_3$, we have
\eqn\xtwobrane{ x_2 = \tr (V_1 V_3) = m^2 + m^{-2} = a^2 -2 }

One can also construct the unknot using identical branes $\tilde \CB$
and the braid group action on one of them:

\ifig\unknotata{Unknot as a union of two branes $\tilde \CB$ with a half-twist.
Each vertical line represents a surface (topologically a 2-sphere) which
divides $\S^3$ into two balls and meets the surface operator at four points.}
{\epsfxsize1.8in\epsfbox{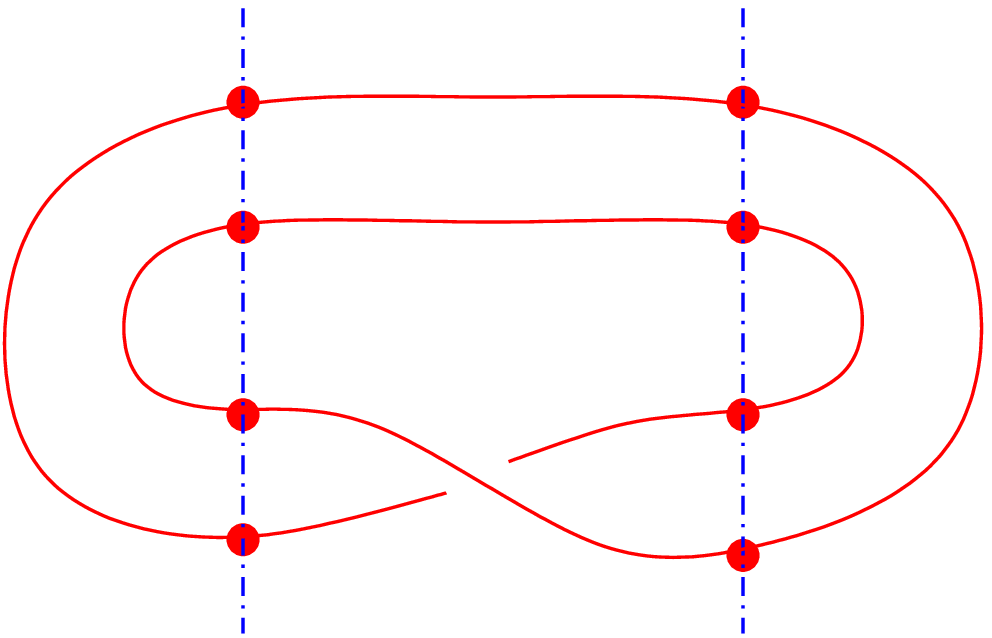}}

\noindent
Here, the two parts of the unknot correspond to the branes
$\tilde \CB$ and $\phi_{\sigma_1} (\tilde \CB)$, where $\tilde \CB$
is the brane described in \albrane\ -- \albranequadric, and
$\sigma_1$ denotes the generator of the braid group $Br_3$.
Using the explicit form \broncubic\ of $\sigma_1$, we find that
the brane $\phi_{\sigma_1} (\tilde \CB)$ is supported on the line:
\eqn\tal{ \phi_{\sigma_1} (\tilde \CB)~:\quad x_2=2}
Together with eq. \xonebrane, this condition implies that the branes
$\tilde \CB$ and $\phi_{\sigma_1} (\tilde \CB)$ meet only at one point
(of multiplicity 2):
\eqn\xxxunknotata{ (x_1,x_2,x_3) = (2,2,a^2-2) }
which is precisely one of the $A_1$ singularities \xxxaone\
in the moduli space $\CM_H$. This is in complete agreement
with our previous analysis, where the same configuration
of D-branes in $\CM$ was found starting from the presentation
of the unknot shown on \unknotaa. This agreement was expected,
of course, since both presentations of the unknot on \unknotaa\
and \unknotata\ are homotopy equivalent in $\S^3$.
The second presentation (on \unknotata) can be easily
generalized to the trefoil knot and more general torus
knots (links) of type $(2,k)$.

\medskip\noindent
$\underline{{\rm Trefoil~Knot:}}$ The trefoil can be constructed
by joining together the brane \albrane\ and the brane
obtained by action of three half-twists on $\tilde \CB$.

\ifig\trefoilattta{The trefoil knot in $\S^3$ can be represented
as a union of two branes $\tilde \CB$ with three half-twists.}
{\epsfxsize1.8in\epsfbox{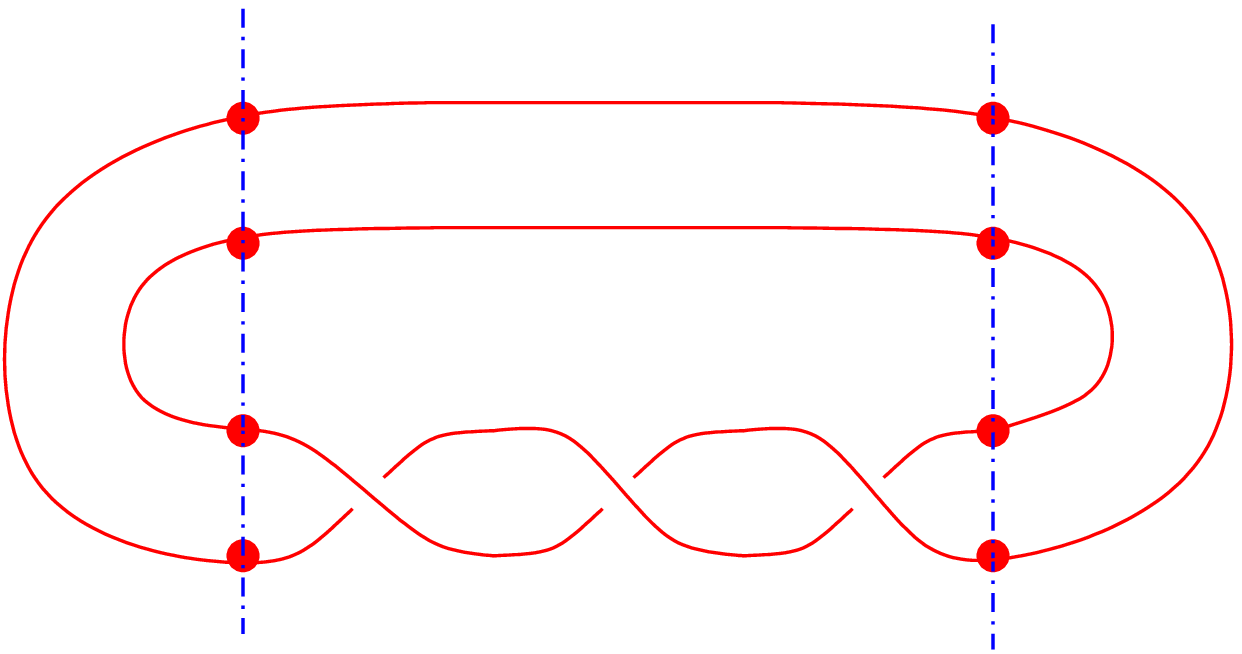}}

Starting with the equation \xonebrane\ descrining the brane $\tilde \CB$
and applying $\sigma_1$ three times, we find that the brane
$\phi_{(\sigma_1)^3} (\tilde \CB)$ is supported on the set of points
\eqn\tttal{ 
(x_1,x_2,x_3) = (4z - 2 a^2 z + 2 a^2 z^2 - 2 z^3 + y (1 - z^2),
-2 + 2 a^2 - 2 a^2 z + y z + 2 z^2, z) }
where we assumed \aaaaa.
Together with the equation $f(x_i)=0$, this condition describes
a subvariety in $\CM_H$ of complex dimension 1.
Using \xonebrane\ and \tttal, it is easy to see that
the branes $\tilde \CB$ and $\phi_{(\sigma_1)^3} (\tilde \CB)$
intersect at two points. The first intersection point
(of multiplicity 2) is precisely the singular point \xxxunknotata,
as in the case of the unknot. The second intersection point
(of multiplicity 4) is located at the regular point in $\CM_H$,
\eqn\xxxtrefoil{ (x_1,x_2,x_3) = (2,a^2-1,1) }
Combining the contributions from the two intersection
points, we find that the cohomology for the trefoil knot
has the following structure
\eqn\htrefoil{
\CH^{sl(2)}_{{\rm trefoil}} = \CH^{sl(2)}_{{\rm unknot}} \oplus \CH^{sl(2)}_{\times} }
where $\CH^{sl(2)}_{{\rm unknot}}$ is the contribution from
the first intersection point, and $\CH^{sl(2)}_{\times}$
denotes the contribution from the new intersection point \xxxtrefoil.
Discarding the contribution of reducible connections,
we find the {\it reduced} cohomology of the trefoil knot,
which consists only of the term $\CH^{sl(2)}_{\times}$,
%
\eqn\htrefcross{\CH^{sl(2)}_{\times} = \IC^4}
Indeed, since $\CM_H$ is smooth near the intersection point,
the configuration of branes $\tilde \CB$ and $\phi_{(\sigma_1)^3} (\tilde \CB)$
can be locally described (in complex structure $J$)
as an intersection of two sets of B-branes in $\C^2$,
such that each set is supported on a line in $\C^2$.
Let us consider a slightly more general problem
where two sets of B-branes in $\C^2$ contain
$n_1$ and $n_2$ branes, respectively.
We denote by $\CE_1$ and $\CE_2$ the corresponding sheaves,
where $\CE_1$ (and similarly $\CE_2$) is defined
by a module of the form $\IC [x_1,x_2]/(x^{n_1})$.
The space of open string states between two such B-branes
is given by
\eqn\extccc{ {\rm Ext}^*_{{\C^2}} (\CE_1,\CE_2) = \IC^{n_1 n_2} }
which, of course, is the expected result since
in the present case open strings form a hypermultiplet
transforming in $(n_1,n_2)$ under $U(n_1) \times U(n_2)$.
Setting $n_1 = n_2 = 2$ gives \htrefcross.


\medskip\noindent
$\underline{(2,k){\rm ~Torus~Knots:}}$
A more general torus knot (link) $T_{2,k}$ can be represented
as a union of two branes $\tilde \CB$ with $k$ half-twists.

\ifig\torusbraidfig{The $(2,k)$ torus knot (link) in $\S^3$ can be represented
as a union of two branes $\tilde \CB$ with $k$ half-twists.}
{\epsfxsize2.2in\epsfbox{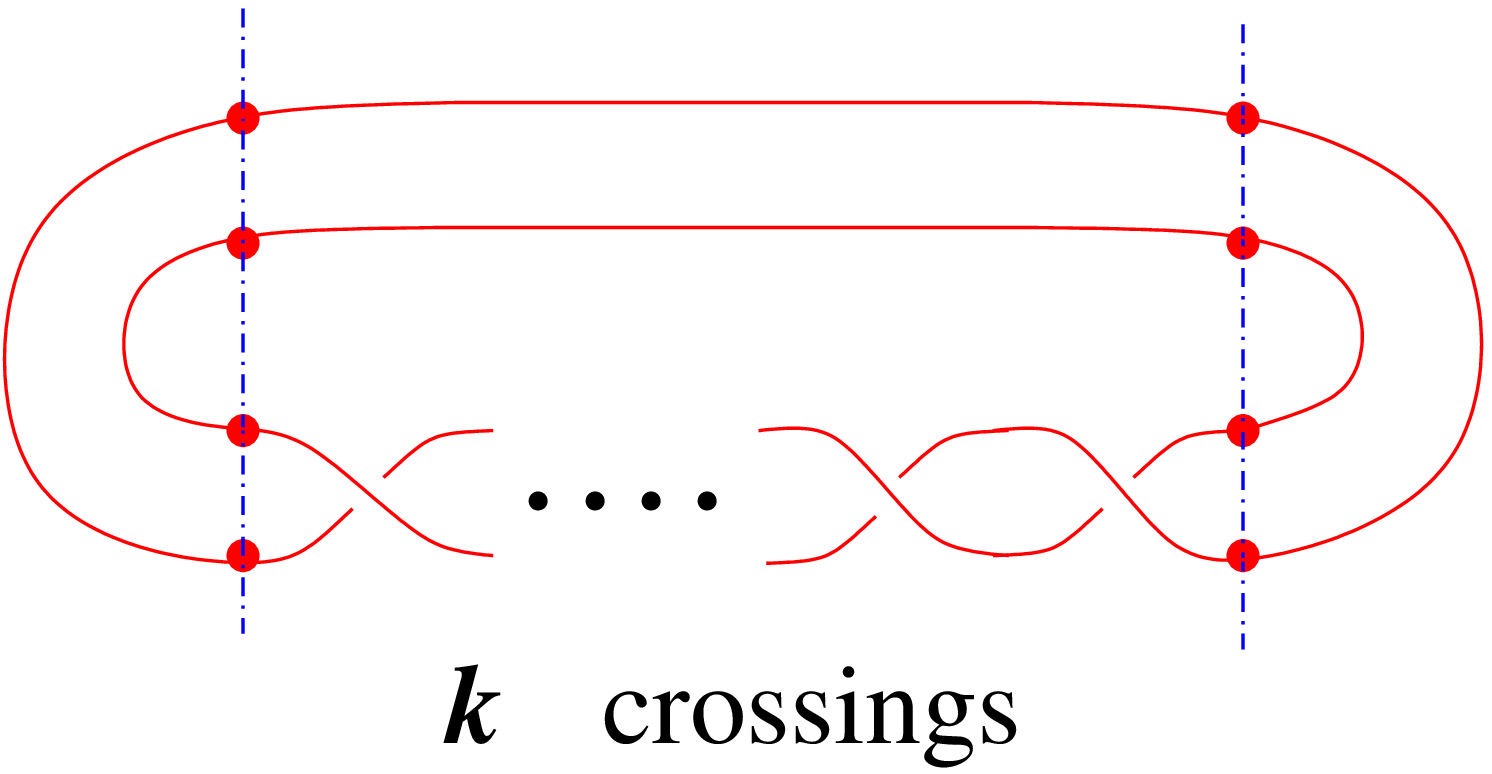}}

In order to describe the action of $(\sigma_1)^k$ on
the brane \albrane, again we use \broncubic.
If the original brane $\tilde \CB$ is represented by a set
of two coincident branes on the line ({\it cf.} \albranequadric),
\eqn\xonebraneline{ (x_1,x_2,x_3) = (2,y,a^2-y) }
the result of $(\sigma_1)^k$ action is a set of branes supported
on a higher-degree curve
\eqn\ttwokababrane{
\phi_{(\sigma_1)^k} (\tilde \CB)~:\quad (x_1,x_2,x_3) = (P_{k}(y) ,P_{k-1}(y) ,a^2 - y) }
where $\{ P_i (y)\}_{i \ge -1}$ is a sequence of polynomials in $y$,
such that $P_0 (y) = 2$, $P_{-1} (y) = y$,
and $P_i (y)$, $i>1$ are determined by the recursion relation
$$
P_{i} (y) = 2a^2 - (a^2-y) P_{i-1} (y) - P_{i-2} (y)
$$
For example, the first few polynomials $P_i (y)$ look like
$$
\eqalign{
& P_1(y) = y \cr
& P_2 (y) = -2 + 2 a^2 - a^2 y + y^2 \cr
& P_3 (y) = 4 a^2 - 2 a^4 - 3 y + 2 a^2 y + a^4 y - 2 a^2 y^2 + y^3 \cr
& P_4 (y) = 2 - 4 a^4 + 2 a^6 + 8 a^2 y - 4 a^4 y - a^6 y - 4 y^2
+ 2 a^2 y^2 + 3 a^4 y^2 - 3 a^2 y^3 + y^4 \cr
& ~~~~~ \vdots }
$$
For simplicity, let us focus on torus knots, which correspond to odd
values of $k$ (the case of $k$ even, which corresponds to torus links,
can be treated similarly). Then, it is easy to see that the brane
$\phi_{(\sigma_1)^k} (\tilde \CB)$ meets the brane \xonebrane\
at $(k+1)/2$ points in $\CM_H$. As in the case of the trefoil knot,
one intersection point (of multiplicity 2) is the point \xxxunknotata\
where $\CM_H$ has $A_1$ singularity due to reducible connections.
The other $(k-1)/2$ points (each of multiplicity 4) are generically
located at regular points in $\CM_H$; their precise location is
determined by the explicit form of $P_i (y)$.
Therefore, extending the earlier result \htrefoil,
we find that cohomology $\CH^{sl(2)}_{T_{2,k}}$
of the torus knot $T_{2,k}$ is isomorphic to a direct
sum of $\CH^{sl(2)}_{{\rm unknot}}$
and $(k-1)/2$ copies of $\CH^{sl(2)}_{\times} = \IC^4$.
As usual, it is convenient to remove the contribution of
reducible solutions. If we denote by $\tilde \CH^{sl(2)}_{K}$
the ``reduced'' cohomology of $K$ for the theory considered here,
we can state our conclusion as
\eqn\hdimtwok{ \dim \tilde \CH^{sl(2)}_{T_{2,k}} = 2 (k-1) }
In general, the cohomology $\tilde \CH^{sl(2)}_{K}$
categorifies a variant of the Casson invariant
obtained by counting flat $SL(2,\IC)$ connections on
the knot complement $\S^3 \setminus K$ with fixed conjugacy
class of the holonomy around the meridian,
\eqn\eulerhsign{ \chi (\tilde \CH^{sl(2)}_{K}) = 2 \sigma (K) }
We expect that, at least for a certain class of knots,
$\sigma (K)$ is the ordinary knot signature.
Notice, that for $(2,k)$ torus knots, we have $\sigma (T_{2,k}) = (k-1)$.

Finally, we note that one could obtain a different knot invariant
(and, presumably, a different knot homology)
by considering the image of the representation variety of
the knot complement in the representation variety of the boundary torus, see {\it e.g.} \WLi.
Indeed, the boundary of the knot complement $Y \setminus K$ can be
identified with $T^2$ in the usual way, and the inclusion
$T^2 \hookrightarrow Y \setminus K$ induces the restriction map
\eqn\rmap{ r : ~\CM_{flat}^{\CCG} (Y \setminus K) \to \CM_{flat}^{\CCG} (T^2) }
which maps a representation $\rho : \pi_1 (Y \setminus K) \to \CCG$
to its restriction $\rho_{\vert T^2} : \pi_1 (T^2) \to \CCG$.
In general, $\CM_{flat}^{\CCG} (Y \setminus K)$ is a branched
cover of its image in $\CM_{flat}^{\CCG} (T^2)$ under the restriction map \rmap.
For example, if $\CCG = SL(2,\IC)$ and $Y = \S^3$ then,
ignoring the multiplicity, the image of the representation variety
$\CM_{flat}^{\CCG} (\S^3 \setminus K)$ under the restriction map
can be described as the zero locus\foot{A-polynomial plays an
important role in quantization of Chern-Simons theory with complex
gauge group $\CCG = SL(2,\IC)$, see \Apol.} of the A-polynomial \CCGLS,
\eqn\zeroofa{ A (l,m) = 0 }
where the complex variables $l$ and $m$ parameterize, respectively,
the conjugacy classes of the holonomy of the flat $SL(2,\IC)$ connection
along the longitude and the meridian of the knot. The A-polynomial of
every knot has a factor $(l-1)$ due to reducible representations.
For example, the A-polynomial of a $(2,k)$ torus knot looks like
\eqn\apoltorus{ A(T_{2,k}) = (l-1)(lm^{2k} + 1) }
Notice, in this example, the part containing irreducible representations
consists of a single curve, $lm^{2k} + 1=0$, of degree one in $l$.
On the other hand, the $SL(2,\IC)$ representation variety of $T_{2,k}$
is a cover of this curve by ${k-1 \over 2}$ distinct irreducible components
which correspond to irreducible representations
counted by $\CN=4$ topological gauge theory.
Restricting the complex variables $l$ and $m$ to be on a unit circle,
we obtain the image of the $SU(2)$ representation variety.
For $(2,k)$ torus knots, the $SU(2)$ representation variety
(again, ignoring reducible representations) is a disjoint
union of ${k-1 \over 2}$ nested open acrs \refs{\Klassen,\BHKK}.


\vskip 30pt

\centerline{\bf Acknowledgments}

\noindent
I would like to thank N.~Dunfield, T.~Hausel, A.~Kapustin,
M.~Khovanov, M.~Mari\~no, T.~Mrowka, J.~Roberts, and C~Vafa
for clarifying discussions and comments.
I am specially indebted to E.~Witten for many useful discussions
and for his observations on a preliminary version of this paper.
This work is supported in part by DOE grant DE-FG03-92-ER40701,
in part by RFBR grant 07-02-00645,
and in part by the grant for support of scientific schools NSh-8004.2006.2.

This paper is an extended version of the talk delivered
at the International Congress on Mathematical Physics 2006 (Rio de Janeiro) and
at the RTN worshop ``Constituents, Fundamental Forces and Symmetries of the Universe'' (Napoli).
I am very grateful to the organizers for the opportunity to participate in these meetings
and to all the participants for providing a stimulating environment.

\listrefs
\end